\documentclass{amsart}
\usepackage{graphicx}
\usepackage{epic,eepic}
\usepackage[normalem]{ulem}% Pour distinguer les contribs
\usepackage{latexsym,amssymb,amsmath,amssymb,xspace}
\newtheorem{thm}{Theorem}[section]
\newtheorem{cor}[thm]{Corollary}
\newtheorem{lem}[thm]{Lemma}
\newtheorem{prop}[thm]{Proposition}
\newtheorem{prob}[thm]{Problem}
\newtheorem{conj}[thm]{Conjecture}
\newtheorem{Clm}{Claim}[thm]

\theoremstyle{definition}

\theoremstyle{remark}

\numberwithin{equation}{section}
\newenvironment{prf}{{\bf \noindent Proof } }{\hfill$\square$\\}
\newenvironment{prfsketch}{{\bf \noindent Sketch of the proof } }{\hfill$\square$\\}
\newenvironment{PrfClaim}{{\bf Proof }}{{\hfill\tiny{$\square$\\}}}

\begin{document}
\input{epsf.sty}
\title[]{On the perfect matching index of bridgeless cubic graphs}
\author{J.L. Fouquet}
\author{J.M. Vanherpe}
\address{L.I.F.O., Facult\'e des Sciences, B.P. 6759
Universit\'e d'Orl\'eans.\newline \indent 45067 Orl\'eans Cedex 2, FR}
%\email{}%

%\thanks{}%
\subjclass{035 C}
\keywords{Cubic graph;  Perfect Matchings;}
%\date
%\dedicatory{}%
%\commby{}%
% ----------------------------------------------------------------
\begin{abstract}
If $G$ is a bridgeless cubic graph, Fulkerson conjectured that we
can find  $6$ perfect matchings $M_1,\ldots,M_6$ of $G$ with the
property that every edge of $G$ is contained in exactly two of them
and Berge conjectured that its edge set can be covered by $5$
perfect matchings. We define $\tau(G)$ as the least number of
perfect matchings allowing to cover the edge set of a bridgeless
cubic graph and we study this parameter. The set of graphs with
perfect matching index $4$ seems interesting and we give some
informations on this class.

\end{abstract}

\maketitle

% ----------------------------------------------------------------
\section{Introduction}
The following conjecture is due to Fulkerson, and appears first in
\cite{Ful71}.
\begin{conj}\label{Conjecture:Fulkerson} If $G$ is a bridgeless
cubic graph, then there exist $6$ perfect matchings $M_1,\ldots,M_6$
of $G$ with the property that every edge of $G$ is contained in
exactly two of $M_1,\ldots,M_6$.
\end{conj}

If $ G $ is 3-edge-colourable, then we may choose three perfect
matchings $ M_1,M_2,M_3 $ so that every edge is in exactly one.
Taking each of these twice gives us $6$ perfect matchings with the
properties described above. Thus, the above conjecture holds
trivially for $3$-edge-colorable graphs. There do exist bridgeless
cubic graphs which are not $3-$edge-colourable (for instance the
Petersen graph), but the above conjecture asserts that every such
graph is close to being $3$-edge-colourable.

If Fulkerson's conjecture were true, then deleting one of the
perfect matchings from the double cover would result in a covering
of the graph by $5$ perfect matchings. This weaker conjecture was
proposed by Berge (see Seymour \cite{Sey79}).

\begin{conj}\label{Conjecture:BergeFulkerson} If $G$ is a bridgeless
cubic graph, then there exists a covering of its edges by  $5$
perfect matchings.
\end{conj}

Since the Petersen graph does not admit a covering by less that $5$
perfect matchings (see section
\ref{Section:PerfectMatchingNumber4}), $5$ in the above conjecture
can not be changed into $4$ and the following weakening of
conjecture \ref{Conjecture:BergeFulkerson} (suggested by Berge) is
still open.

\begin{conj}\label{Conjecture:Berge1} There exists a fixed integer $ k $ such that
the edge set of every bridgeless cubic graph can be written as a
union of $ k $ perfect matchings.
\end{conj}

Another consequence of the Fulkerson conjecture would be that every
bridgeless cubic graph has $3$ perfect matchings with empty
intersection (take any $3$ of the $6$ perfect matchings given by the
conjecture). The following weakening of this (also suggested by
Berge) is still open.

\begin{conj}\label{Conjecture:Berge2}There exists a fixed integer $ k $
such that every bridgeless cubic graph has a list of $ k $ perfect
matchings with empty intersection.
\end{conj}

For $k=3$ this conjecture is known as the Fan Raspaud Conjecture.

\begin{conj}\cite{FanRas} \label{Conjecture:FanRaspaud} Every
bridgeless cubic graph contains perfect matching $M_1$, $M_2$, $M_3$
such that
$$M_1 \cap M_2 \cap M_3 = \emptyset$$
\end{conj}

While some partial results exist concerning conjecture
\ref{Conjecture:FanRaspaud} (see \cite{HaoEtal09}), we have noticed no result in
the literature concerning the validity of Conjecture
\ref{Conjecture:Fulkerson} or Conjecture \ref{Conjecture:Berge2} for
te usual classes of graphs which are examined when dealing with the
$5-$flow conjecture of Tutte \cite{Tut54} or the cycle double
conjecture of Seymour \cite{Sey79b} and Szekeres \cite{Sze73}. Hence
for bridgeless cubic graphs with {\em oddness} $2$ (a $2-$factor
contains exactly tow odd cycles) it is known that the $5-$flow
conjecture holds true as well as the cycle double conjecture (see
Zhang \cite{Zha97} for a comprehensive study of this subject).

Let $G$ be a bridgeless cubic graph, we shall say that the set
$\mathcal M =\{M_{1}, \ldots M_{k}\}$ ($k \geq 3$) of perfect
matchings is a {\em $k-$covering} when each edge is contained  in at
least one of theses perfect matchings. A {\em Fulkerson covering} is a $6-$covering where each edge appears
exactly twice. Since every edge of a bridgeless cubic graph is contained in a perfect matching (see
\cite{Sch34}) the minimum number $\tau(G)$ of perfect matchings
covering its edge set is well defined. We shall say that $\tau(G)$
is the {\em perfect matching index} of $G$. We obviously have that
$\tau(G)=3$ if and only if $G$ is $3-$edge-colourable.

%\begin{figure}\label{Figure:Composition}
%\centering
%\input{J5p.eepic}
%\caption{Composition of two cubic graphs} \label{Figure:Composition}
%\end{figure}
%%
%\begin{figure}[htb]
%\centering \epsfsize=0.4 \hsize \noindent \epsfbox{cprp.eps}
%\caption{$CPR_{p}$} \label{Figure:CPRp}
%\end{figure}
%
\section{\label{Section:Preliminaries}Preliminaries results}

\begin{prop}\label{Proposition:k_elementaire} let $G$ be a cubic graph with  a $k-$covering
$\mathcal M =\{M_{1}, \ldots,M_{k}\}$ ($k \geq 3$) then  $G$ is
bridgeless.
\end{prop}
\begin{prf}
Assume that $e \in E(G)$ is an isthmus, then the edges incident to
$e$ are not covered by any perfect matching of $G$ and $\mathcal M$
is not a $K-$covering, a contradiction.
\end{prf}

%\begin{prop}\label{Proposition:5_elementaire} let $G$ be a cubic graph with  a $5-$covering
%$\mathcal M =\{M_{1}, \ldots,M_{5}\}$ ($k \geq 3$) such that each
%edge appears at most twice then the set of edges contained in
%exactly one perfect matching is a perfect matching.
%\end{prop}
%\begin{prf}
%Let $v$ be any vertex of $G$, each edge incident with $v$ must be
%contained in some perfect matching of $\mathcal M$ and  each perfect
%matching must be incident with $v$. We have thus exactly one edge
%incident with $v$ which is covered by exactly one perfect matching
%of $\mathcal M$ while the two other edges are covered by exactly two
%perfect matchings. The result follows.
%\end{prf}

\subsection{$2-$cut connection}
Let $G_{1}$, $G_{2}$ be two bridgeless cubic graph and
$e_{1}=u_{1}v_{1} \in E(G_{1})$, $e_{2}=u_{2}v_{2} \in E(G_{1})$ be
two edges. Construct a new graph $G=G_{1}\bigodot G_{2}$
$$G=[G_{1}
\setminus \{e_{1}\}] \cup [G_{2} \setminus \{e_{2}\}] \cup
\{u_{1}u_{2},v_{1}v_{2}\}$$

\begin{prop}\label{Proposition:Construction:2Cut} Let $G_{1}$ be a cubic graph such that $\tau(G_{1})=k \geq 3$
and let $G_{2}$ be any cubic bridgeless graph, then
$\tau(G_{1}\bigodot G_{2}) \geq k $
\end{prop}
\begin{prf}
Let $G=G_{1}\bigodot G_{2}$. Assume that $k'=\tau(G) <  k $ and let $\mathcal M =\{M_{1}, \ldots
,M_{k'}\}$ be a $k'-$covering of $G$  Any perfect matching of $G$
must intersect the $2-$edge cut $\{u_{1}v_{1},u_{2}v_{2}\}$ in two
edges or has no edge in common with that set. Thus any perfect
matching in $\mathcal M$ leads to a perfect matching of $G_{1}$.
Hence we should have a $k'-$covering of the edge set of $G_{1}$, a
contradiction.
\end{prf}

\subsection{$3-$cut connection}
Let $G_{1}$, $G_{2}$ be two bridgeless cubic graph and $u \in
V(G_{1})$, $v \in V(G_{2})$ be two vertices with
$N(u)=\{u_{1},u_{2},u_{3}\}$  and $N(v)=\{v_{1},v_{2},v_{3}\}$.
Construct a new graph $G=G_{1}\otimes G_{2}$
$$G=[G_{1} \setminus \{u\}] \cup [G_{2} \setminus \{v\}] \cup \{u_{1}v_{1},u_{2}v_{2},u_{3}v_{3}\}$$
It is well known that the resulting graph $G_{1}\otimes G_{2}$ is
bridgeless. The $3-$edge cut $\{u_{1}v_{1},u_{2}v_{2},u_{3}v_{3}\}$
will be called the {\em principal $3-$edge cut}.

\begin{prop}\label{Proposition:Construction:3OddCut} Let $G_{1}$ be a cubic graph such that $\tau(G_{1})=k \geq 3$
and let $G_{2}$ be any cubic bridgeless graph. Let
$k'=\tau(G_{1}\otimes G_{2})$ and let $\mathcal M =\{M_{1}, \ldots
,M_{k'}\}$ be a $k'-$covering of $G_{1}\otimes G_{2}$. Then one of
the followings is true
\begin{enumerate}
  \item \label{Item:Construction:3OddCutOne} $k' \geq k$
  \item \label{Item:Construction:3OddCutTwo}  There is a perfect
  matching $M_{i} \in \mathcal M$ ($1 \leq i \leq k$) containing the principal $3-$edge cut

\end{enumerate}
\end{prop}
\begin{prf}
Assume that $k'< k$. Any perfect matching of $G_{1}\otimes G_{2}$
must intersect the principal $3-$edge cut in one or three edges. If
none of the perfect matchings in $\mathcal M$ contains the principal
$3-$edge cut, then any perfect matching in  $\mathcal M$ leads to a
perfect matching of $G_{1}$ and any edge of $G_{1}$ is covered by
one of these perfect matchings. Hence we should have a $k'-$covering
of the edge set of $G_{1}$, a contradiction.
\end{prf}

\section{\label{Section:PerfectMatchingNumber4} On graphs with perfect matching index $4$}

A natural question is to investigate the class of graphs for which
the perfect matching index is $4$.

\begin{prop}\label{Proposition:Tau4Elementaire} Let $G$ be a cubic graph with  a $4-$covering
$\mathcal M =\{M_{1}, M_{2},M_{3},M_{4}\}$ then
\begin{enumerate}
  \item \label{Item:TwoOne} Every edge is contained in exactly one or two perfect matchings of $\mathcal M$.
  \item \label{Item:TwoPerfectMatching}  The set $M$ of edges contained in exactly two perfect matchings of $\mathcal M$ is a  perfect matching.
  \item \label{Item:MatchingsNonDisjoints} If $\tau(G)=4$ then $\forall i \neq j  \in \{1,2,3,4\} \quad M_{i} \cap M_{j} \neq \emptyset$.
\end{enumerate}
\end{prop}
\begin{prf}
Let $v$ be any vertex of $G$, each
edge incident with $v$ must be contained in some perfect matching of
$\mathcal M$ and  each perfect matching must be incident with $v$.
We have thus exactly one edge incident with $v$ which is covered by
exactly two perfect matchings of $\mathcal M$ while the two other
edges are covered by exactly one perfect matching. We get thus
immediately Items \ref{Item:TwoOne} and
\ref{Item:TwoPerfectMatching}.

When  $\tau(G)=4$, $G$ is not a $3-$edge colourable graph. Assume
that we have two perfect matchings with an
empty intersection. These two perfect matchings lead to an even
$2-$factor and hence a a $3-$edge colouring of $G$, a contradiction.
\end{prf}
In the following the edges of the matching $M$ described in item \ref{Item:TwoPerfectMatching} of Proposition \ref{Proposition:Tau4Elementaire} will be said to be {\em covered twice}.
\begin{prop} \label{Proposition:12vertices} Let $G$ be a cubic graph such that $\tau(G)=4$ then $G$
has at least $12$ vertices
\end{prop}
\begin{prf}
Let $\mathcal M =\{M_{1}, M_{2},M_{3},M_{4}\}$ be a covering of the
edge set of $G$ into $4$ perfect matchings. From Proposition
\ref{Proposition:Tau4Elementaire} we must have at least $6$ edges
in the perfect matching formed with the edges covered twice in $\mathcal M$. Hence, $G$ must have at least $12$ vertices as claimed.
\end{prf}

From Proposition \ref{Proposition:12vertices}, we obviously have
that the Petersen graph has a perfect matching index equal to $5$.

\begin{prop} \label{Proposition:4FR} Let $G$ be a cubic graph such that $\tau(G)=4$
and let $\mathcal M =\{M_{1}, M_{2},M_{3},M_{4}\}$ be a covering of
its edge set into $4$ perfect matchings then for each $j$ ($j=1
\ldots 4$) $\mathcal M -M_{j}$ is a set of $3$ perfect matchings
satisfying the Fan Raspaud conjecture.
\end{prop}
\begin{prf}
Obvious since, by Item \ref{Item:TwoOne} of Proposition
\ref{Proposition:Tau4Elementaire} any edge is contained in exactly
one or two perfect matchings of $\mathcal M$.
\end{prf}

Let $G$ be a cubic graph with $3$ perfect matchings $M_{1}, M_{2}$
and $M_{3}$ having an empty intersection. Since such a graph satisfy
the Fan Raspaud conjecture, when considering these three perfect
matchings, we shall say that $(M_{1}, M_{2},M_{3})$ is an {\em
FR-triple}. When a cubic graph has a FR-triple we define $T_i$ ($i=0,1,2$) as the set of edges that belong to precisely $i$ matchings of the FR-triple. Thus $(T_0, T_1, T_2)$ is a partition of the edge set.

\begin{prop} \label{Proposition:StructureFR}
Let $G$ be a cubic graph with $3$ perfect matchings $M_{1}, M_{2}$
and $M_{3}$ having an empty intersection. Then the set $T_0\cup T_2$ is a set of disjoint even cycles. Moreover, the  edges of $T_0$ and $T_2$ alternate along these cycles.
\end{prop}
\begin{prf}
Let $v$ be a vertex incident to a edge of $T_0$. Since $v$ must be
incident to each perfect matching and since the three perfect
matchings have an empty intersection, one of the remaining edges
incident to $v$ must be contained into $2$ perfect matchings while
the other is contained in exactly one perfect matching. The result
follows.
\end{prf}

Let $G$ be a bridgeless cubic graph and let $C$ and $C'$ be distinct odd cycles of $G$. Assume that there are three distinct
edges namely $xx'$, $yy'$ and $zz'$ such that $x$, $y$ and $z$ are vertices of $C$ while $x'$, $y'$, $z'$ are vertices of $C'$ which
determine on $C$ and on $C'$ edge-disjoint paths of odd length then we shall say that $(xx', yy', zz')$ is a {\em good triple} and that
the pair of cycles $\{C,C'\}$ is  a {\em good pair}.

\begin{thm} \label{Theorem:2Factor_GoodPair}
Let $G$ be a cubic graph which has a $2-$factor $F$ whose odd cycles
can be arranged into good pairs $\{C_{1},D_{1}\}$,$\{C_{2},D_{2}\}$,
..., $\{C_{k},D_{k}\}$. Then $\tau(G) \leq 4$.
\end{thm}
\begin{prf}
For each good pair $\{C_{i},D_{i}\}$ let
$(c^{1}_{i}d^{1}_{i},c^{2}_{i}d^{2}_{i},c^{3}_{i}d^{3}_{i})$ be a good triple of $C_{i}$ and $D_{i}$, $c^{1}_{i}$, $c^{2}_{i}$, $c^{3}_{i}$ being vertices of $C_{i}$ while $d^{1}_{i}$, $d^{2}_{i}$ and  $^{3}_{i}$ are on $D_{i}$. In order to construct a set $\mathcal M
=\{M_{1}, M_{2},M_{3},M_{4}\}$ of $4$ perfect matchings covering the
edge set of $G$ we let $M_{1}$ as the perfect matching of $G$
obtained by deleting the edges of the $2-$factor.

Let $A_{j}$ be the set of edges $\{c^{j}_{i}d^{j}_{i} | i=1 \ldots
k\}$. We construct a perfect matching $M_{j}$ ($j=2,3,4$) of $G$
such that $M_{1} \cap M_{j} = A_{j}$. For each good pair
$\{C_{i},D_{i}\}$ ($i=1 \ldots k$), we add to $A_{j}$ the unique
perfect matching
 contained in $E(C_{i}) \cup E(D_{i})$ when the two vertices
 $c^{j}_{i}$ and $d^{j}_{i}$ are deleted. We get hence $3$ matchings
 $B_{j}$ ($j=2,3,4$) where each vertex contained in a good pair is
 saturated.
 If the $2-$factor contains some even cycles, we add first a perfect
 matching contained in the edge set of these even cycles to $B_{2}$.
 We obtain thus a perfect matching $M_{2}$ whose intersection with
 $M_{1}$ is reduced to $A_{2}$. The remaining edges of these even
 cycles are added to $B_{3}$ and to $B_{4}$, leading to the perfect
 matchings $M_{3}$ and $M_{4}$. Let us remark that each edge of
 these even cycles are contained in  $M_{2} \cup M_{3}$.

We claim that each edge of $G$ is contained in at least one of
$\mathcal M =\{M_{1}, M_{2},M_{3},M_{4}\}$. Since $M_{1}$ is the
perfect matching which complements in $G$ the $2-$factor $F$,  the
above remark says that we have just to prove that each edge of each
good pair is covered by some perfect matching of $\mathcal M$. By
construction, no edge is contained in $M_{1} \cup M_{2} \cup M_{3}$
which means that $(M_{1}, M_{2},M_{3})$ is an FR-triple. In the same
way, $(M_{1}, M_{3},M_{4})$ and $(M_{1}, M_{2},M_{4})$ are
FR-triples. The edges of $T_0\cup T_2$ induced by the FR-triple $(M_{1}, M_{2}, M_{3})$ on each good pair $\{C_{i},D_{i}\}$ is the even cycle
$\Gamma_{i}$  using $c^{1}_{i}d^{1}_{i}$ and $c^{2}_{i}d^{2}_{i}$,
the odd path of $C_{i}$ joining $c^{1}_{i}$ to $c^{2}_{i}$ and the
odd path of $D_{i}$ joining $d^{1}_{i}$ to $d^{2}_{i}$. In the same
way, edges of $T_0\cup T_2$ induced by the FR-triple $(M_{1}, M_{3}, M_{4})$ on each good pair $\{C_{i},D_{i}\}$ is the even cycle
$\Lambda_{i}$ using $c^{2}_{i}d^{2}_{i}$ and $c^{3}_{i}d^{3}_{i}$,
the odd path of $C_{i}$ joining $c^{2}_{i}$ to $c^{3}_{i}$ and the
odd path of $D_{i}$ joining $d^{2}_{i}$ to $d^{3}_{i}$. It is an
easy task to see that these two cycles $\Gamma_{i}$ and
$\Lambda_{i}$  have the only edge $c^{2}_{i}d^{2}_{i}$ in common.
Hence each edge of $\Gamma_{i}\cap T_0$ is contained into $M_4$ while
each edge of $\Lambda_{i}\cap T_0$  is contained into $M_{2}$. The result follows.
\end{prf}

\subsection{\label{Subsection:BalancedMatching} On balanced
matchings}

A set $A \subseteq E(G)$ is a {\em balanced matching} when we can
find $2$ perfect matchings $M_1$ and $M_2$ such that $A=M_1 \cap
M_2$. Let $B(G)$ be the set of balanced matchings of $G$, we define
$b(G)$ as the minimum size of a any set $ A \in B(G)$, we have:

\begin{prop}\label{Proposition:BalancematchingAuPlusnsur12} Let $G$ be a cubic graph such that $\tau(G)=4$
then $b(G) \leq \frac{n}{12}$.
\end{prop}
\begin{prf}Let
$\mathcal M =\{M_{1}, M_{2},M_{3},M_{4}\}$ be a covering of the edge
set of $G$ into $4$ perfect matchings and let $M$ be the perfect
matching of edges contained in exactly two perfect
  matchings of $\mathcal M$ (Iem \ref{Item:TwoPerfectMatching} of
  Proposition \ref{Proposition:4_elementaire}). Since $M_{i} \cap M_{j} \neq
  \emptyset$ $\forall i \neq j  \in
\{1,2,3,4\}$ by Proposition \ref{Proposition:tau4_elementaire},
these $6$ balanced matchings partition $M$. Hence, one of them must
have at most $\frac{|M|}{6}=\frac{n}{12}$ edges.

\end{prf}

In \cite{KaiKraNor} Kaiser, Kr\'{a}l and Norine proved
\begin{thm} \label{Theorem:KaiserKralNorine} Any bridgeless cubic
graph contains $2$ perfect matchings whose union cover at least
$\frac{9n}{10}$ edges of $G$.
\end{thm}

From Theorem \ref{Theorem:KaiserKralNorine}, we can find two perfect
matchings with an intersection having at most $\frac{n}{10}$ edges
in any cubic bridgeless graph.  It can be proved (see
\cite{FouVan08} ) that for any cyclically $4$-edge connected cubic
graph $G$, either $b(G) \leq \frac{n}{14}$ or any perfect matching
contains an odd cut of size $5$.

\subsection{\label{SubSection:PerfectMatchingNumber4:BalnusaIsaacGoldberg} On classical snarks}
\begin{figure}[htb]
%\centering \epsfsize=0.6 \hsize \noindent \epsfbox{Blanusa_1.eps}
\begin{center}
\setlength{\unitlength}{0.00026247in}
\begingroup\makeatletter\ifx\SetFigFont\undefined
% extract first six characters in \fmtname
\def\x#1#2#3#4#5#6#7\relax{\def\x{#1#2#3#4#5#6}}%
\expandafter\x\fmtname xxxxxx\relax \def\y{splain}%
\ifx\x\y   % LaTeX or SliTeX?
\gdef\SetFigFont#1#2#3{%
  \ifnum #1<17\tiny\else \ifnum #1<20\small\else
  \ifnum #1<24\normalsize\else \ifnum #1<29\large\else
  \ifnum #1<34\Large\else \ifnum #1<41\LARGE\else
     \huge\fi\fi\fi\fi\fi\fi
  \csname #3\endcsname}%
\else
\gdef\SetFigFont#1#2#3{\begingroup
  \count@#1\relax \ifnum 25<\count@\count@25\fi
  \def\x{\endgroup\@setsize\SetFigFont{#2pt}}%
  \expandafter\x
    \csname \romannumeral\the\count@ pt\expandafter\endcsname
    \csname @\romannumeral\the\count@ pt\endcsname
  \csname #3\endcsname}%
\fi
\fi\endgroup
{\renewcommand{\dashlinestretch}{30}
\begin{picture}(12675,5965)(0,-10)
\put(4217,5134){\makebox(0,0)[lb]{\smash{{{\SetFigFont{6}{7.2}{rm}x}}}}}
\texture{ffffffff ffeeeeee eeffffff fffbfbfb fbffffff ffeeeeee eeffffff ffbfbbbf 
	bbffffff ffeeeeee eeffffff fffbfbfb fbffffff ffeeeeee eeffffff ffbfbfbf 
	bfffffff ffeeeeee eeffffff fffbfbfb fbffffff ffeeeeee eeffffff ffbfbbbf 
	bbffffff ffeeeeee eeffffff fffbfbfb fbffffff ffeeeeee eeffffff ffbfbfbf }
\put(12499,1336){\shade\ellipse{336}{336}}
\put(12499,1336){\ellipse{336}{336}}
\put(12499,4612){\shade\ellipse{336}{336}}
\put(12499,4612){\ellipse{336}{336}}
\put(4221,3050){\shade\ellipse{336}{336}}
\put(4221,3050){\ellipse{336}{336}}
\put(4221,1366){\shade\ellipse{336}{336}}
\put(4221,1366){\ellipse{336}{336}}
\put(4221,4642){\shade\ellipse{336}{336}}
\put(4221,4642){\ellipse{336}{336}}
\thicklines
\put(6313.000,-10528.000){\arc{32863.319}{4.3294}{5.1014}}
\put(6313.000,16477.000){\arc{32862.604}{1.1818}{1.9538}}
\thinlines
\put(10467,4630){\shade\ellipse{336}{336}}
\put(10467,4630){\ellipse{336}{336}}
\put(10467,1318){\shade\ellipse{336}{336}}
\put(10467,1318){\ellipse{336}{336}}
\put(8454,3020){\shade\ellipse{336}{336}}
\put(8454,3020){\ellipse{336}{336}}
\put(8472,1300){\shade\ellipse{336}{336}}
\put(8472,1300){\ellipse{336}{336}}
\put(8454,4649){\shade\ellipse{336}{336}}
\put(8454,4649){\ellipse{336}{336}}
\put(6221,1354){\shade\ellipse{336}{336}}
\put(6221,1354){\ellipse{336}{336}}
\put(6221,4630){\shade\ellipse{336}{336}}
\put(6221,4630){\ellipse{336}{336}}
\put(2189,4660){\shade\ellipse{336}{336}}
\put(2189,4660){\ellipse{336}{336}}
\put(2189,1348){\shade\ellipse{336}{336}}
\put(2189,1348){\ellipse{336}{336}}
\put(176,3050){\shade\ellipse{336}{336}}
\put(176,3050){\ellipse{336}{336}}
\put(194,1330){\shade\ellipse{336}{336}}
\put(194,1330){\ellipse{336}{336}}
\put(176,4679){\shade\ellipse{336}{336}}
\put(176,4679){\ellipse{336}{336}}
\thicklines
\dashline{150.000}(8445,4695)(10550,4695)
\dashline{150.000}(6194,4695)(6194,1327)
\dashline{150.000}(2186,4695)(4272,4695)
\path(173,1364)(2168,1382)
\path(191,4713)(191,3011)
\path(10476,1346)(12563,1346)
\path(12490,4695)(12490,2938)
\path(8463,3084)(8463,1309)
\path(4217,3084)(4217,1401)
\path(173,4713)(2259,4713)
\path(2168,1382)(4199,1382)
\path(2186,4713)(2186,1272)
\path(10464,4683)(10464,1242)
\path(4254,1370)(8500,1351)
\path(8500,1327)(10513,1327)
\path(10495,4676)(12599,4676)
\path(12508,3048)(12508,1327)
\path(8445,4695)(8445,3029)
\path(8451,3054)(12532,3054)
\path(173,3084)(4254,3084)
\path(173,3084)(173,1327)
\path(4217,4695)(4217,3066)
\path(4236,4695)(8482,4676)
\put(10385,4988){\makebox(0,0)[lb]{\smash{{{\SetFigFont{6}{7.2}{rm}z'}}}}}
\put(6176,797){\makebox(0,0)[lb]{\smash{{{\SetFigFont{6}{7.2}{rm}y'}}}}}
\put(2168,5024){\makebox(0,0)[lb]{\smash{{{\SetFigFont{6}{7.2}{rm}x'}}}}}
\put(8408,5042){\makebox(0,0)[lb]{\smash{{{\SetFigFont{6}{7.2}{rm}z}}}}}
\put(6176,5079){\makebox(0,0)[lb]{\smash{{{\SetFigFont{6}{7.2}{rm}y}}}}}
\thinlines
\put(12499,3020){\shade\ellipse{336}{336}}
\put(12499,3020){\ellipse{336}{336}}
\end{picture}
}
\end{center}
\caption{ Blanu\v{s}a snark \#1} \label{Figure:Blanusa_1}
\end{figure}
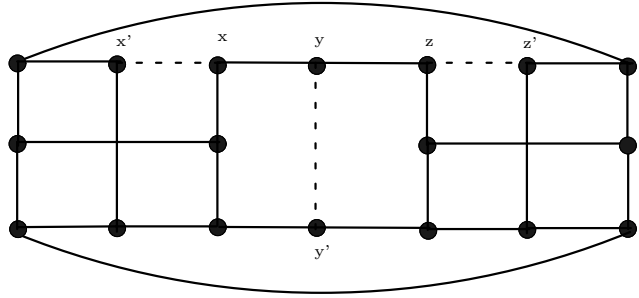

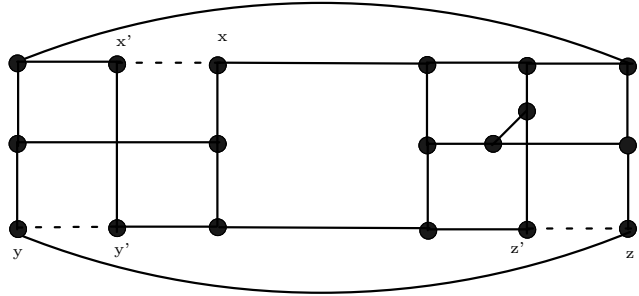
\begin{figure}[htb]
%\centering \epsfsize=0.6 \hsize \noindent \epsfbox{blanusa_2.eps}
\begin{center}
\setlength{\unitlength}{0.00026247in}
\begingroup\makeatletter\ifx\SetFigFont\undefined
% extract first six characters in \fmtname
\def\x#1#2#3#4#5#6#7\relax{\def\x{#1#2#3#4#5#6}}%
\expandafter\x\fmtname xxxxxx\relax \def\y{splain}%
\ifx\x\y   % LaTeX or SliTeX?
\gdef\SetFigFont#1#2#3{%
  \ifnum #1<17\tiny\else \ifnum #1<20\small\else
  \ifnum #1<24\normalsize\else \ifnum #1<29\large\else
  \ifnum #1<34\Large\else \ifnum #1<41\LARGE\else
     \huge\fi\fi\fi\fi\fi\fi
  \csname #3\endcsname}%
\else
\gdef\SetFigFont#1#2#3{\begingroup
  \count@#1\relax \ifnum 25<\count@\count@25\fi
  \def\x{\endgroup\@setsize\SetFigFont{#2pt}}%
  \expandafter\x
    \csname \romannumeral\the\count@ pt\expandafter\endcsname
    \csname @\romannumeral\the\count@ pt\endcsname
  \csname #3\endcsname}%
\fi
\fi\endgroup
{\renewcommand{\dashlinestretch}{30}
\begin{picture}(12675,5965)(0,-10)
\put(4217,5134){\makebox(0,0)[lb]{\smash{{{\SetFigFont{6}{7.2}{rm}x}}}}}
\texture{ffffffff ffeeeeee eeffffff fffbfbfb fbffffff ffeeeeee eeffffff ffbfbbbf 
	bbffffff ffeeeeee eeffffff fffbfbfb fbffffff ffeeeeee eeffffff ffbfbfbf 
	bfffffff ffeeeeee eeffffff fffbfbfb fbffffff ffeeeeee eeffffff ffbfbbbf 
	bbffffff ffeeeeee eeffffff fffbfbfb fbffffff ffeeeeee eeffffff ffbfbfbf }
\put(12499,1336){\shade\ellipse{336}{336}}
\put(12499,1336){\ellipse{336}{336}}
\put(12499,4612){\shade\ellipse{336}{336}}
\put(12499,4612){\ellipse{336}{336}}
\put(4221,3050){\shade\ellipse{336}{336}}
\put(4221,3050){\ellipse{336}{336}}
\put(4221,1366){\shade\ellipse{336}{336}}
\put(4221,1366){\ellipse{336}{336}}
\put(4221,4642){\shade\ellipse{336}{336}}
\put(4221,4642){\ellipse{336}{336}}
\thicklines
\put(6313.000,-10528.000){\arc{32863.319}{4.3294}{5.1014}}
\put(6313.000,16477.000){\arc{32862.604}{1.1818}{1.9538}}
\thinlines
\put(10461,3709){\shade\ellipse{336}{336}}
\put(10461,3709){\ellipse{336}{336}}
\put(9784,3050){\shade\ellipse{336}{336}}
\put(9784,3050){\ellipse{336}{336}}
\put(10467,4630){\shade\ellipse{336}{336}}
\put(10467,4630){\ellipse{336}{336}}
\put(10467,1318){\shade\ellipse{336}{336}}
\put(10467,1318){\ellipse{336}{336}}
\put(8454,3020){\shade\ellipse{336}{336}}
\put(8454,3020){\ellipse{336}{336}}
\put(8472,1300){\shade\ellipse{336}{336}}
\put(8472,1300){\ellipse{336}{336}}
\put(8454,4649){\shade\ellipse{336}{336}}
\put(8454,4649){\ellipse{336}{336}}
\put(2189,4660){\shade\ellipse{336}{336}}
\put(2189,4660){\ellipse{336}{336}}
\put(2189,1348){\shade\ellipse{336}{336}}
\put(2189,1348){\ellipse{336}{336}}
\put(176,3050){\shade\ellipse{336}{336}}
\put(176,3050){\ellipse{336}{336}}
\put(194,1330){\shade\ellipse{336}{336}}
\put(194,1330){\ellipse{336}{336}}
\put(176,4679){\shade\ellipse{336}{336}}
\put(176,4679){\ellipse{336}{336}}
\thicklines
\dashline{150.000}(173,1364)(2168,1382)
\dashline{150.000}(10476,1346)(12563,1346)
\path(8445,4695)(10550,4695)
\path(10458,3725)(9763,3029)
\dashline{150.000}(2186,4695)(4272,4695)
\path(191,4713)(191,3011)
\path(12490,4695)(12490,2938)
\path(8463,3084)(8463,1309)
\path(4217,3084)(4217,1401)
\path(173,4713)(2259,4713)
\path(2168,1382)(4199,1382)
\path(2186,4713)(2186,1272)
\path(10464,4683)(10464,1242)
\path(4254,1370)(8500,1351)
\path(8500,1327)(10513,1327)
\path(10495,4676)(12599,4676)
\path(12508,3048)(12508,1327)
\path(8445,4695)(8445,3029)
\path(8451,3054)(12532,3054)
\path(173,3084)(4254,3084)
\path(173,3084)(173,1327)
\path(4217,4695)(4217,3066)
\path(4236,4695)(8482,4676)
\put(12453,742){\makebox(0,0)[lb]{\smash{{{\SetFigFont{6}{7.2}{rm}z}}}}}
\put(10129,851){\makebox(0,0)[lb]{\smash{{{\SetFigFont{6}{7.2}{rm}z'}}}}}
\put(81,797){\makebox(0,0)[lb]{\smash{{{\SetFigFont{6}{7.2}{rm}y}}}}}
\put(2131,833){\makebox(0,0)[lb]{\smash{{{\SetFigFont{6}{7.2}{rm}y'}}}}}
\put(2168,5024){\makebox(0,0)[lb]{\smash{{{\SetFigFont{6}{7.2}{rm}x'}}}}}
\thinlines
\put(12499,3020){\shade\ellipse{336}{336}}
\put(12499,3020){\ellipse{336}{336}}
\end{picture}
}
\end{center}
\caption{ Blanu\v{s}a snark \#2} \label{Figure:Blanusa_2}
\end{figure}

As usual a {\em snark} is a non $3-$edge colourable bridgeless cubic
graph.
In Figure \ref{Figure:Blanusa_1} is depicted one of the two the
Blanu\v{s}a snarks on $18$ vertices \cite{Bla46}. In bold we have
drawn a $2-$factor (each cycle has length $9$) and the dashed edges
connect the triple $(x,y,z)$ of one cycle  to the triple
$(x',y',z')$ of the second cycle. It is a routine matter to check
that $(xx', yy', zz')$ is a good triple and Theorem
\ref{Theorem:2Factor_GoodPair} allows us to say that this graph has
perfect matching index $4$. In the same way the second Blanu\v{s}a
snark on $18$ vertices depicted in Figure \ref{Figure:Blanusa_2} can
be covered by $4$ perfect matchings by using Theorem
\ref{Theorem:2Factor_GoodPair}.

For an odd $k\geq 3$ the Flower Snark $F_k$ intoduced by Isaac (see
\cite{Isa75}) is the cubic graph on $4k$ vertices $x_0,x_1,\ldots
x_{k-1}$, $y_0,y_1,\ldots y_{k-1}$, $z_0,z_1,\ldots z_{k-1}$,
$t_0,t_1,\ldots t_{k-1}$ such that $x_0x_1\ldots x_{k-1}$ is an
induced cycle of length $k$, $y_0y_1\ldots y_{k-1}$ $z_0z_1\ldots
z_{k-1}$ is an induced cycle of length $2k$ and for $i=0\ldots k-1$
the vertex $t_i$ is adjacent to $x_i$, $y_i$ and $z_i$. The set
$\{t_i,x_i,y_i,z_i\}$ induces the claw $C_i$. In Figure
\ref{Figure:F5} we have a representation of $F_{5}$, the half edges
(to the left and to the right in the figure) with same labels are
identified.

\begin{figure}[htb]
%\centering \epsfsize=0.6 \hsize \noindent \epsfbox{j5.eps}
\begin{center}
\setlength{\unitlength}{0.00026247in}
\begingroup\makeatletter\ifx\SetFigFont\undefined
% extract first six characters in \fmtname
\def\x#1#2#3#4#5#6#7\relax{\def\x{#1#2#3#4#5#6}}%
\expandafter\x\fmtname xxxxxx\relax \def\y{splain}%
\ifx\x\y   % LaTeX or SliTeX?
\gdef\SetFigFont#1#2#3{%
  \ifnum #1<17\tiny\else \ifnum #1<20\small\else
  \ifnum #1<24\normalsize\else \ifnum #1<29\large\else
  \ifnum #1<34\Large\else \ifnum #1<41\LARGE\else
     \huge\fi\fi\fi\fi\fi\fi
  \csname #3\endcsname}%
\else
\gdef\SetFigFont#1#2#3{\begingroup
  \count@#1\relax \ifnum 25<\count@\count@25\fi
  \def\x{\endgroup\@setsize\SetFigFont{#2pt}}%
  \expandafter\x
    \csname \romannumeral\the\count@ pt\expandafter\endcsname
    \csname @\romannumeral\the\count@ pt\endcsname
  \csname #3\endcsname}%
\fi
\fi\endgroup
{\renewcommand{\dashlinestretch}{30}
\begin{picture}(16974,4735)(0,-10)
\put(140,4237){\makebox(0,0)[lb]{\smash{{{\SetFigFont{6}{7.2}{rm}a}}}}}
\put(15643,3894){\blacken\ellipse{482}{482}}
\put(15643,3894){\ellipse{482}{482}}
\put(13806,3206){\blacken\ellipse{482}{482}}
\put(13806,3206){\ellipse{482}{482}}
\put(14681,2031){\blacken\ellipse{482}{482}}
\put(14681,2031){\ellipse{482}{482}}
\put(11587,594){\blacken\ellipse{482}{482}}
\put(11587,594){\ellipse{482}{482}}
\put(12549,3894){\blacken\ellipse{482}{482}}
\put(12549,3894){\ellipse{482}{482}}
\put(10712,3206){\blacken\ellipse{482}{482}}
\put(10712,3206){\ellipse{482}{482}}
\put(11587,2031){\blacken\ellipse{482}{482}}
\put(11587,2031){\ellipse{482}{482}}
\put(8391,594){\blacken\ellipse{482}{482}}
\put(8391,594){\ellipse{482}{482}}
\put(9353,3894){\blacken\ellipse{482}{482}}
\put(9353,3894){\ellipse{482}{482}}
\put(7516,3206){\blacken\ellipse{482}{482}}
\put(7516,3206){\ellipse{482}{482}}
\put(8391,2031){\blacken\ellipse{482}{482}}
\put(8391,2031){\ellipse{482}{482}}
\put(5203,607){\blacken\ellipse{482}{482}}
\put(5203,607){\ellipse{482}{482}}
\put(6165,3907){\blacken\ellipse{482}{482}}
\put(6165,3907){\ellipse{482}{482}}
\put(4328,3219){\blacken\ellipse{482}{482}}
\put(4328,3219){\ellipse{482}{482}}
\put(5203,2044){\blacken\ellipse{482}{482}}
\put(5203,2044){\ellipse{482}{482}}
\put(2059,607){\blacken\ellipse{482}{482}}
\put(2059,607){\ellipse{482}{482}}
\put(3021,3907){\blacken\ellipse{482}{482}}
\put(3021,3907){\ellipse{482}{482}}
\put(1184,3219){\blacken\ellipse{482}{482}}
\put(1184,3219){\ellipse{482}{482}}
\put(2059,2044){\blacken\ellipse{482}{482}}
\put(2059,2044){\ellipse{482}{482}}
\thicklines
\dashline{150.000}(3034,3944)(2059,2056)
\dashline{150.000}(6178,3944)(5203,2056)
\dashline{150.000}(9366,3931)(8391,2043)
\path(2071,2044)(2059,656)
\path(2071,2044)(2059,656)
\path(2053,650)(5303,650)
\path(5215,2044)(5203,656)
\path(5215,2044)(5203,656)
\path(4328,3256)(5203,2056)
\path(4328,3256)(5203,2056)
\path(7491,3218)(4315,3218)
\path(7491,3218)(4315,3218)
\path(7516,3243)(8391,2043)
\path(7516,3243)(8391,2043)
\path(8403,2031)(8391,643)
\path(8403,2031)(8391,643)
\path(8403,625)(11690,625)
\path(11599,2031)(11587,643)
\path(11599,2031)(11587,643)
\path(10712,3243)(11587,2043)
\path(10712,3243)(11587,2043)
\path(10690,3212)(13803,3212)
\path(10690,3212)(13803,3212)
\path(13806,3243)(14681,2043)
\path(13806,3243)(14681,2043)
\path(14693,2031)(14681,643)
\path(14693,2031)(14681,643)
\path(14697,576)(16909,589)
\path(14697,576)(16909,589)
\path(15622,3900)(16897,3887)
\path(15622,3900)(16897,3887)
\path(12528,3900)(15611,3891)
\path(12528,3900)(15611,3891)
\path(9403,3918)(12278,3912)
\path(9403,3918)(12278,3912)
\path(9391,3906)(6216,3906)
\path(9391,3906)(6216,3906)
\path(3003,3925)(6153,3925)
\path(3028,3925)(65,3937)
\path(1184,3256)(2059,2056)
\path(1184,3256)(2059,2056)
\path(1165,3225)(90,3225)
\path(15656,3931)(14681,2043)
\path(15656,3931)(14681,2043)
\path(12562,3931)(11587,2043)
\path(12562,3931)(11587,2043)
\path(11603,602)(14687,610)
\path(11603,602)(14687,610)
\path(13784,3212)(16897,3212)
\path(13784,3212)(16897,3212)
\path(7540,3225)(10728,3225)
\path(1178,3225)(4365,3225)
\path(2053,637)(115,637)
\path(8353,618)(5178,618)
\path(8353,618)(5178,618)
\put(14603,12){\makebox(0,0)[lb]{\smash{{{\SetFigFont{6}{7.2}{rm}z4}}}}}
\put(15003,2050){\makebox(0,0)[lb]{\smash{{{\SetFigFont{6}{7.2}{rm}t4}}}}}
\put(13603,2625){\makebox(0,0)[lb]{\smash{{{\SetFigFont{6}{7.2}{rm}y4}}}}}
\put(11515,0){\makebox(0,0)[lb]{\smash{{{\SetFigFont{6}{7.2}{rm}z3}}}}}
\put(11903,2050){\makebox(0,0)[lb]{\smash{{{\SetFigFont{6}{7.2}{rm}t3}}}}}
\put(10515,2612){\makebox(0,0)[lb]{\smash{{{\SetFigFont{6}{7.2}{rm}y3}}}}}
\put(1990,12){\makebox(0,0)[lb]{\smash{{{\SetFigFont{6}{7.2}{rm}z0}}}}}
\put(2465,2037){\makebox(0,0)[lb]{\smash{{{\SetFigFont{6}{7.2}{rm}t0}}}}}
\put(978,2650){\makebox(0,0)[lb]{\smash{{{\SetFigFont{6}{7.2}{rm}y0}}}}}
\put(5115,25){\makebox(0,0)[lb]{\smash{{{\SetFigFont{6}{7.2}{rm}z1}}}}}
\put(5553,2075){\makebox(0,0)[lb]{\smash{{{\SetFigFont{6}{7.2}{rm}t1}}}}}
\put(4128,2650){\makebox(0,0)[lb]{\smash{{{\SetFigFont{6}{7.2}{rm}y1}}}}}
\put(8315,12){\makebox(0,0)[lb]{\smash{{{\SetFigFont{6}{7.2}{rm}z2}}}}}
\put(7340,2587){\makebox(0,0)[lb]{\smash{{{\SetFigFont{6}{7.2}{rm}y2}}}}}
\put(8740,2075){\makebox(0,0)[lb]{\smash{{{\SetFigFont{6}{7.2}{rm}t2}}}}}
\put(9303,4162){\makebox(0,0)[lb]{\smash{{{\SetFigFont{6}{7.2}{rm}x2}}}}}
\put(6140,4200){\makebox(0,0)[lb]{\smash{{{\SetFigFont{6}{7.2}{rm}x1}}}}}
\put(2965,4287){\makebox(0,0)[lb]{\smash{{{\SetFigFont{6}{7.2}{rm}x0}}}}}
\put(16778,650){\makebox(0,0)[lb]{\smash{{{\SetFigFont{6}{7.2}{rm}b}}}}}
\put(16790,3262){\makebox(0,0)[lb]{\smash{{{\SetFigFont{6}{7.2}{rm}c}}}}}
\put(16803,3962){\makebox(0,0)[lb]{\smash{{{\SetFigFont{6}{7.2}{rm}a}}}}}
\put(115,712){\makebox(0,0)[lb]{\smash{{{\SetFigFont{6}{7.2}{rm}c}}}}}
\put(103,3350){\makebox(0,0)[lb]{\smash{{{\SetFigFont{6}{7.2}{rm}b}}}}}
\thinlines
\put(14681,594){\blacken\ellipse{482}{482}}
\put(14681,594){\ellipse{482}{482}}
\end{picture}
}
\end{center}
\caption{$J_{5}$} \label{Figure:F5}
\end{figure}

\begin{thm}  $\tau(F_{k})=4$.
\end{thm}
\begin{prf} Let $k=2p+1 \geq 3$ and let  $C=x_{0}x_{1} \ldots x_{2p}$
, $D=y_{0}t{0}z_{0}z_{1}t_{1}y_{1}$ $\ldots$
$y_{2i}t_{2i}z_{2i}z_{2i+1}t_{2i+1}y_{2i+1}$ $\ldots$
$y_{2p}t_{2p}z_{2p}$ ($0 \leq i \leq$) be the odd cycles of lengths
$2k+1$ and $3 \times (2k+1)$ respectively which partition $F_{k}$
(in bold in Figure \ref{Figure:F5}. It is a routine matter
to check that the edges $x_0t_0$, $x_1t_1$ and $x_2t_2$ form a good
triple (dashed edges in Figure \ref{Figure:F5}). Hence $(C,D)$ is a
$2-$factor of $G$ and it is a good pair. The result follows from
Theorem \ref{Theorem:2Factor_GoodPair}.
\end{prf}

Let $H$ be the graph depicted in Figure \ref{Figure:Hi}

\begin{figure}[htb]
%\centering \epsfsize=0.15 \hsize \noindent \epsfbox{Hi.eps}
\begin{center}
\setlength{\unitlength}{0.00017498in}
\begingroup\makeatletter\ifx\SetFigFont\undefined
% extract first six characters in \fmtname
\def\x#1#2#3#4#5#6#7\relax{\def\x{#1#2#3#4#5#6}}%
\expandafter\x\fmtname xxxxxx\relax \def\y{splain}%
\ifx\x\y   % LaTeX or SliTeX?
\gdef\SetFigFont#1#2#3{%
  \ifnum #1<17\tiny\else \ifnum #1<20\small\else
  \ifnum #1<24\normalsize\else \ifnum #1<29\large\else
  \ifnum #1<34\Large\else \ifnum #1<41\LARGE\else
     \huge\fi\fi\fi\fi\fi\fi
  \csname #3\endcsname}%
\else
\gdef\SetFigFont#1#2#3{\begingroup
  \count@#1\relax \ifnum 25<\count@\count@25\fi
  \def\x{\endgroup\@setsize\SetFigFont{#2pt}}%
  \expandafter\x
    \csname \romannumeral\the\count@ pt\expandafter\endcsname
    \csname @\romannumeral\the\count@ pt\endcsname
  \csname #3\endcsname}%
\fi
\fi\endgroup
{\renewcommand{\dashlinestretch}{30}
\begin{picture}(4388,7418)(0,-10)
\put(1525,6803){\makebox(0,0)[lb]{\smash{{{\SetFigFont{6}{7.2}{rm}a}}}}}
\put(2374,5480){\blacken\ellipse{482}{482}}
\put(2374,5480){\ellipse{482}{482}}
\put(782,3796){\blacken\ellipse{482}{482}}
\put(782,3796){\ellipse{482}{482}}
\put(3710,3778){\blacken\ellipse{482}{482}}
\put(3710,3778){\ellipse{482}{482}}
\put(3171,249){\blacken\ellipse{482}{482}}
\put(3171,249){\ellipse{482}{482}}
\put(3171,1686){\blacken\ellipse{482}{482}}
\put(3171,1686){\ellipse{482}{482}}
\put(1323,286){\blacken\ellipse{482}{482}}
\put(1323,286){\ellipse{482}{482}}
\put(1323,1723){\blacken\ellipse{482}{482}}
\put(1323,1723){\ellipse{482}{482}}
\path(800,3796)(3765,3796)
\path(1349,245)(3179,245)
\path(2374,5480)(2374,6852)
\path(2337,5480)(1294,1710)
\path(2392,5480)(3179,1600)
\path(1294,1728)(3710,3778)
\path(782,3778)(3198,1636)
\thicklines
\path(3191,1649)(3179,261)
\path(3191,1649)(3179,261)
\path(1335,1723)(1331,410)
\path(1335,1723)(1331,410)
\put(0,3753){\makebox(0,0)[lb]{\smash{{{\SetFigFont{6}{7.2}{rm}c}}}}}
\put(4150,3740){\makebox(0,0)[lb]{\smash{{{\SetFigFont{6}{7.2}{rm}h}}}}}
\put(3963,1603){\makebox(0,0)[lb]{\smash{{{\SetFigFont{6}{7.2}{rm}g}}}}}
\put(3963,190){\makebox(0,0)[lb]{\smash{{{\SetFigFont{6}{7.2}{rm}f}}}}}
\put(100,190){\makebox(0,0)[lb]{\smash{{{\SetFigFont{6}{7.2}{rm}e}}}}}
\put(125,1590){\makebox(0,0)[lb]{\smash{{{\SetFigFont{6}{7.2}{rm}d}}}}}
\put(1513,5365){\makebox(0,0)[lb]{\smash{{{\SetFigFont{6}{7.2}{rm}b}}}}}
\thinlines
\put(2374,6852){\blacken\ellipse{482}{482}}
\put(2374,6852){\ellipse{482}{482}}
\end{picture}
}
\end{center}
\caption{$H$} \label{Figure:Hi}
\end{figure}

Let $G_{k}$ ($k$ odd) be a cubic graph obtained from $k$ copies of
$H$ ($H_{0} \ldots H_{k-1}$ where the name of vertices are indexed
by $i$) in adding edges  $a_{i}a_{i+1}$, $c_{i}c_{i+1}$,
$e_{i}e_{i+1}$, $f_{i}f_{i+1}$ and $h_{i}h_{i+1}$ (subscripts are
taken modulo $k$).

\begin{figure}[htb]
%\centering \epsfsize=0.9 \hsize \noindent \epsfbox{g5.eps}
\begin{center}
\setlength{\unitlength}{0.00017498in}
\begingroup\makeatletter\ifx\SetFigFont\undefined
% extract first six characters in \fmtname
\def\x#1#2#3#4#5#6#7\relax{\def\x{#1#2#3#4#5#6}}%
\expandafter\x\fmtname xxxxxx\relax \def\y{splain}%
\ifx\x\y   % LaTeX or SliTeX?
\gdef\SetFigFont#1#2#3{%
  \ifnum #1<17\tiny\else \ifnum #1<20\small\else
  \ifnum #1<24\normalsize\else \ifnum #1<29\large\else
  \ifnum #1<34\Large\else \ifnum #1<41\LARGE\else
     \huge\fi\fi\fi\fi\fi\fi
  \csname #3\endcsname}%
\else
\gdef\SetFigFont#1#2#3{\begingroup
  \count@#1\relax \ifnum 25<\count@\count@25\fi
  \def\x{\endgroup\@setsize\SetFigFont{#2pt}}%
  \expandafter\x
    \csname \romannumeral\the\count@ pt\expandafter\endcsname
    \csname @\romannumeral\the\count@ pt\endcsname
  \csname #3\endcsname}%
\fi
\fi\endgroup
{\renewcommand{\dashlinestretch}{30}
\begin{picture}(25595,7364)(0,-10)
\put(87,6638){\makebox(0,0)[lb]{\smash{{{\SetFigFont{6}{7.2}{rm}x}}}}}
\put(12794,5655){\blacken\ellipse{482}{482}}
\put(12794,5655){\ellipse{482}{482}}
\put(11202,3971){\blacken\ellipse{482}{482}}
\put(11202,3971){\ellipse{482}{482}}
\put(14130,3953){\blacken\ellipse{482}{482}}
\put(14130,3953){\ellipse{482}{482}}
\put(13591,424){\blacken\ellipse{482}{482}}
\put(13591,424){\ellipse{482}{482}}
\put(13591,1861){\blacken\ellipse{482}{482}}
\put(13591,1861){\ellipse{482}{482}}
\put(11743,461){\blacken\ellipse{482}{482}}
\put(11743,461){\ellipse{482}{482}}
\put(11743,1898){\blacken\ellipse{482}{482}}
\put(11743,1898){\ellipse{482}{482}}
\put(17660,7045){\blacken\ellipse{482}{482}}
\put(17660,7045){\ellipse{482}{482}}
\put(17660,5673){\blacken\ellipse{482}{482}}
\put(17660,5673){\ellipse{482}{482}}
\put(16068,3989){\blacken\ellipse{482}{482}}
\put(16068,3989){\ellipse{482}{482}}
\put(18996,3971){\blacken\ellipse{482}{482}}
\put(18996,3971){\ellipse{482}{482}}
\put(18457,442){\blacken\ellipse{482}{482}}
\put(18457,442){\ellipse{482}{482}}
\put(18457,1879){\blacken\ellipse{482}{482}}
\put(18457,1879){\ellipse{482}{482}}
\put(16609,479){\blacken\ellipse{482}{482}}
\put(16609,479){\ellipse{482}{482}}
\put(16609,1916){\blacken\ellipse{482}{482}}
\put(16609,1916){\ellipse{482}{482}}
\put(22433,7018){\blacken\ellipse{482}{482}}
\put(22433,7018){\ellipse{482}{482}}
\put(22433,5646){\blacken\ellipse{482}{482}}
\put(22433,5646){\ellipse{482}{482}}
\put(20841,3962){\blacken\ellipse{482}{482}}
\put(20841,3962){\ellipse{482}{482}}
\put(23769,3944){\blacken\ellipse{482}{482}}
\put(23769,3944){\ellipse{482}{482}}
\put(23230,415){\blacken\ellipse{482}{482}}
\put(23230,415){\ellipse{482}{482}}
\put(23230,1852){\blacken\ellipse{482}{482}}
\put(23230,1852){\ellipse{482}{482}}
\put(21382,452){\blacken\ellipse{482}{482}}
\put(21382,452){\ellipse{482}{482}}
\put(21382,1889){\blacken\ellipse{482}{482}}
\put(21382,1889){\ellipse{482}{482}}
\put(8017,7045){\blacken\ellipse{482}{482}}
\put(8017,7045){\ellipse{482}{482}}
\put(8017,5673){\blacken\ellipse{482}{482}}
\put(8017,5673){\ellipse{482}{482}}
\put(6425,3989){\blacken\ellipse{482}{482}}
\put(6425,3989){\ellipse{482}{482}}
\put(9353,3971){\blacken\ellipse{482}{482}}
\put(9353,3971){\ellipse{482}{482}}
\put(8814,442){\blacken\ellipse{482}{482}}
\put(8814,442){\ellipse{482}{482}}
\put(8814,1879){\blacken\ellipse{482}{482}}
\put(8814,1879){\ellipse{482}{482}}
\put(6966,479){\blacken\ellipse{482}{482}}
\put(6966,479){\ellipse{482}{482}}
\put(6966,1916){\blacken\ellipse{482}{482}}
\put(6966,1916){\ellipse{482}{482}}
\put(3286,7100){\blacken\ellipse{482}{482}}
\put(3286,7100){\ellipse{482}{482}}
\put(3286,5728){\blacken\ellipse{482}{482}}
\put(3286,5728){\ellipse{482}{482}}
\put(1694,4044){\blacken\ellipse{482}{482}}
\put(1694,4044){\ellipse{482}{482}}
\put(4622,4026){\blacken\ellipse{482}{482}}
\put(4622,4026){\ellipse{482}{482}}
\put(4083,497){\blacken\ellipse{482}{482}}
\put(4083,497){\ellipse{482}{482}}
\put(4083,1934){\blacken\ellipse{482}{482}}
\put(4083,1934){\ellipse{482}{482}}
\put(2235,534){\blacken\ellipse{482}{482}}
\put(2235,534){\ellipse{482}{482}}
\put(2235,1971){\blacken\ellipse{482}{482}}
\put(2235,1971){\ellipse{482}{482}}
\path(22433,5646)(22433,7018)
\path(17660,5673)(17660,7045)
\thicklines
\dashline{180.000}(3286,5728)(3286,7100)
\dashline{210.000}(8017,5673)(8017,7045)
\dashline{210.000}(12794,5655)(12794,7027)
\path(23645,3971)(25347,3971)
\path(20859,3998)(23824,3998)
\path(19004,4013)(20907,4013)
\path(16086,4025)(19051,4025)
\path(14152,4016)(16055,4016)
\path(11220,4007)(14185,4007)
\path(9344,4026)(11247,4026)
\path(6443,4025)(9408,4025)
\path(4659,4044)(6342,4044)
\path(1712,4044)(4677,4044)
\path(65,4044)(1767,4044)
\path(102,493)(2206,493)
\path(2247,1971)(2243,658)
\path(2247,1971)(2243,658)
\path(3249,5728)(2206,1958)
\path(3304,5728)(4091,1848)
\path(4103,1897)(4091,509)
\path(4103,1897)(4091,509)
\path(4055,512)(6983,512)
\path(6978,1916)(6974,603)
\path(6978,1916)(6974,603)
\path(7980,5673)(6937,1903)
\path(8035,5673)(8822,1793)
\path(8834,1842)(8822,454)
\path(8834,1842)(8822,454)
\path(8831,493)(11759,493)
\path(11755,1898)(11751,585)
\path(11755,1898)(11751,585)
\path(12757,5655)(11714,1885)
\path(12812,5655)(13599,1775)
\path(13611,1824)(13599,436)
\path(13611,1824)(13599,436)
\path(13577,470)(16505,470)
\path(16621,1916)(16617,603)
\path(16621,1916)(16617,603)
\path(17623,5673)(16580,1903)
\path(17678,5673)(18465,1793)
\path(18477,1842)(18465,454)
\path(18477,1842)(18465,454)
\path(18449,449)(21377,449)
\path(21394,1889)(21390,576)
\path(21394,1889)(21390,576)
\path(22396,5646)(21353,1876)
\path(22451,5646)(23238,1766)
\path(23250,1815)(23238,427)
\path(23250,1815)(23238,427)
\thinlines
\path(11769,456)(13599,456)
\path(11714,1903)(14130,3953)
\path(11202,3953)(13618,1811)
\path(16635,474)(18465,474)
\path(16580,1921)(18996,3971)
\path(16068,3971)(18484,1829)
\path(21408,447)(23238,447)
\path(21353,1894)(23769,3944)
\path(20841,3944)(23257,1802)
\thicklines
\path(23292,439)(25396,439)
\path(22537,7089)(25390,7089)
\path(102,7100)(22486,7093)
\thinlines
\path(6992,474)(8822,474)
\path(6937,1921)(9353,3971)
\path(6425,3971)(8841,1829)
\path(2261,493)(4091,493)
\path(2206,1976)(4622,4026)
\path(1694,4026)(4110,1884)
\put(87,3613){\makebox(0,0)[lb]{\smash{{{\SetFigFont{6}{7.2}{rm}y}}}}}
\put(25390,0){\makebox(0,0)[lb]{\smash{{{\SetFigFont{6}{7.2}{rm}z}}}}}
\put(25366,3508){\makebox(0,0)[lb]{\smash{{{\SetFigFont{6}{7.2}{rm}y}}}}}
\put(25293,6577){\makebox(0,0)[lb]{\smash{{{\SetFigFont{6}{7.2}{rm}x}}}}}
\put(87,13){\makebox(0,0)[lb]{\smash{{{\SetFigFont{6}{7.2}{rm}z}}}}}
\put(12794,7027){\blacken\ellipse{482}{482}}
\put(12794,7027){\ellipse{482}{482}}
\end{picture}
}
\end{center}
\caption{Goldberg snark $G_5$} \label{Figure:GoldbergG5}
\end{figure}
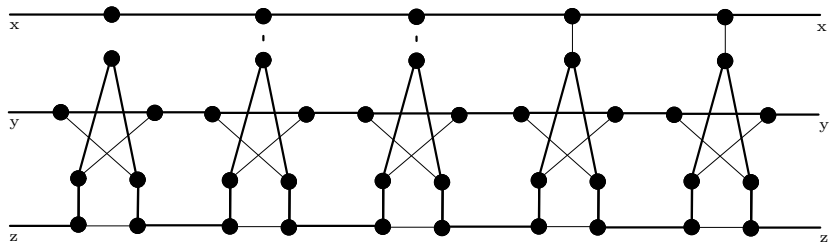

If $k = 5$, then $G_k$ is known as the Goldberg snark. Accordingly,
we refer to all graphs $G_k$ as Goldberg graphs. The graph $G_5$ is
shown in Figure \ref{Figure:GoldbergG5}. The half edges (to the left
and to the right in the figure) with same labels are identified.

\begin{thm}  $\tau(G_{k})=4$.
\end{thm}
\begin{prf} Let $k=2p+1 \geq 3$ and let  $C=a_{0}a_{1} \ldots a_{2p}$
, $D=e_{0}d_{0}b_{0}g_{0}f_{0}$ $e_{1}d_{1}b_{1}g_{1}f_{1}$ $\ldots$
 $e_{i}d_{i}b_{i}g_{i}f_{i}$ $\ldots$ $e_{2p}d_{2p}b_{2p}g_{2p}f_{2p}$
 ($0 \leq i \leq$) be the odd cycles of lengths
$2k+1$ and $5 \times (2k+1)$ respectively and
$E=c_{0}h_{0}c_{1}h_{1}$ $\ldots$ $c_{i}h_{i}$ $\ldots$
$c_{2p}h_{2p}$ the cycle of length $4k$ of $G_{k}$. This set of $3$
cycles is a $2-$factor of $G_{k}$ (in bold in Figure
\ref{Figure:GoldbergG5}). At last, $a_{0}b_{0}$, $a_{1}b_{1}$ and
$a_{2}b_{2}$ are edges of $G$ (dashed edges in Figure
\ref{Figure:GoldbergG5}). Then $(a_0b_0, a_1b_1, a_2b_2)$ is
a good triple. Hence $(C,D,E)$ is a $2-$factor of $G$ where $(C,D)$
is a good pair . The result follows from Theorem
\ref{Theorem:2Factor_GoodPair}.
\end{prf}

%\begin{figure}[htb]
%  % Requires \usepackage{graphicx}
%  \includegraphics[bb=0 0 100 100]{GoldberG5.jpg}\\
%  \caption{Goldberg snark $G_5$}\label{Figure:GoldbergG5}
%\end{figure}
\subsection{\label{SubSection:HomeoPetersen} On  permutation graphs}

A cubic graph $G$ is called a {\em permutation graph} if $G$ has a
$2-$factor $F$ such that $F$ is the union of two chordless cycles
$C$ and $C'$. Let $M$ be the perfect matching $G-F$. A
subgraph homeomorphic to the Petersen graph with no edge of $M$
subdivided is called a $M-P_{10}$. Ellingham \cite{Ell84} showed
that a permutation graph without any $M-P_{10}$ is $3-$edge
colourable.

In general, we do not know whether a permutation graph
distinct from the Petersen graph is $3-$edge colourable or not. It
is an easy task to construct a cyclically $4-$edge connected
permutation graph which is a snark (consider the two Blanusa snarks
on $18$ vertices for exemple) and Zhang \cite{Zha97} conjectured:

\begin{conj}
Let $G$ be a $3-$connected cyclically $5-$edge connected permutation
graph. If $G$ is a snark, then $G$ must be the Petersen graph.
\end{conj}

Let us consider a permutation graph $G$ with a $2$-factor $F$ having two cycles $C$ and $C'$. Two distinct vertices of $C$
say $x$ and $y$ determine on $C$ two paths with $x$ and $y$ as end-points.
In order to be unambiguous when considering those paths from their end-points we give an  orientation to $C$.
Thus $C(x,y)$ will denote in the following the path of $C$ that starts with the vertex $x$ and ends with the vertex $x$ according
to the orientation of $C$.The notation $C'(x',y')$ is defined similarly when $x'$ and $y'$ are vertices of $C'$.

In order to determine which permutation graphs have a perfect
matching index less than $4$ we state the following tool (see
Figure \ref{Fig:C8dansPermutationGraph})~:
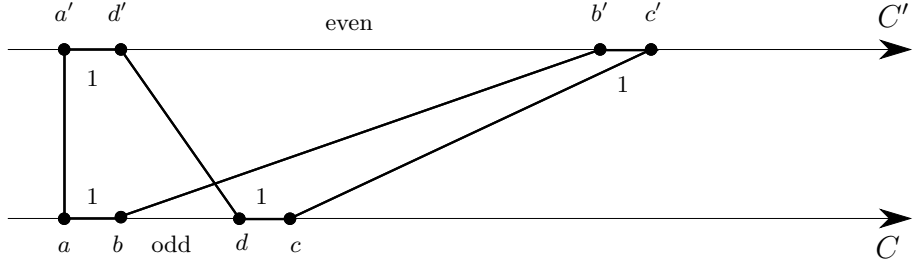
\begin{figure}
\setlength{\unitlength}{0.00065617in}
\begingroup\makeatletter\ifx\SetFigFont\undefined
% extract first six characters in \fmtname
\def\x#1#2#3#4#5#6#7\relax{\def\x{#1#2#3#4#5#6}}%
\expandafter\x\fmtname xxxxxx\relax \def\y{splain}%
\ifx\x\y   % LaTeX or SliTeX?
\gdef\SetFigFont#1#2#3{%
  \ifnum #1<17\tiny\else \ifnum #1<20\small\else
  \ifnum #1<24\normalsize\else \ifnum #1<29\large\else
  \ifnum #1<34\Large\else \ifnum #1<41\LARGE\else
     \huge\fi\fi\fi\fi\fi\fi
  \csname #3\endcsname}%
\else
\gdef\SetFigFont#1#2#3{\begingroup
  \count@#1\relax \ifnum 25<\count@\count@25\fi
  \def\x{\endgroup\@setsize\SetFigFont{#2pt}}%
  \expandafter\x
    \csname \romannumeral\the\count@ pt\expandafter\endcsname
    \csname @\romannumeral\the\count@ pt\endcsname
  \csname #3\endcsname}%
\fi
\fi\endgroup
{\renewcommand{\dashlinestretch}{30}
\begin{picture}(7224,2183)(0,-10)
\put(687,1451){\makebox(0,0)[b]{\smash{{{\SetFigFont{9}{10.8}{rm}$1$}}}}}
\put(912,404){\blacken\ellipse{90}{90}}
\put(912,404){\ellipse{90}{90}}
\put(2262,393){\blacken\ellipse{90}{90}}
\put(2262,393){\ellipse{90}{90}}
\put(1857,393){\blacken\ellipse{90}{90}}
\put(1857,393){\ellipse{90}{90}}
\put(5142,1743){\blacken\ellipse{90}{90}}
\put(5142,1743){\ellipse{90}{90}}
\put(912,1743){\blacken\ellipse{90}{90}}
\put(912,1743){\ellipse{90}{90}}
\put(462,1743){\blacken\ellipse{90}{90}}
\put(462,1743){\ellipse{90}{90}}
\put(462,393){\blacken\ellipse{90}{90}}
\put(462,393){\ellipse{90}{90}}
\thicklines
\path(2262,393)(5142,1743)
\path(2262,393)(5142,1743)
\path(912,1743)(1857,393)
\path(912,1743)(1857,393)
\path(4737,1743)(900,405)
\path(4737,1743)(900,405)
\path(462,1743)(462,393)
\path(462,1743)(462,393)
\path(4742,1738)(5192,1738)
\path(4742,1738)(5192,1738)
\path(1851,388)(2301,388)
\path(1851,388)(2301,388)
\path(462,393)(912,393)
\path(462,393)(912,393)
\path(462,1743)(912,1743)
\path(462,1743)(912,1743)
\thinlines
\path(12,393)(7212,393)
\blacken\path(6972.000,321.000)(7212.000,393.000)(6972.000,465.000)(7044.000,393.000)(6972.000,321.000)
\path(12,1743)(7212,1743)
\blacken\path(6972.000,1671.000)(7212.000,1743.000)(6972.000,1815.000)(7044.000,1743.000)(6972.000,1671.000)
\put(5164,1968){\makebox(0,0)[b]{\smash{{{\SetFigFont{9}{10.8}{rm}$c'$}}}}}
\put(4737,1968){\makebox(0,0)[b]{\smash{{{\SetFigFont{9}{10.8}{rm}$b'$}}}}}
\put(889,1968){\makebox(0,0)[b]{\smash{{{\SetFigFont{9}{10.8}{rm}$d'$}}}}}
\put(462,1968){\makebox(0,0)[b]{\smash{{{\SetFigFont{9}{10.8}{rm}$a'$}}}}}
\put(2307,124){\makebox(0,0)[b]{\smash{{{\SetFigFont{9}{10.8}{rm}$c$}}}}}
\put(1879,146){\makebox(0,0)[b]{\smash{{{\SetFigFont{9}{10.8}{rm}$d$}}}}}
\put(889,124){\makebox(0,0)[b]{\smash{{{\SetFigFont{9}{10.8}{rm}$b$}}}}}
\put(462,124){\makebox(0,0)[b]{\smash{{{\SetFigFont{9}{10.8}{rm}$a$}}}}}
\put(1317,123){\makebox(0,0)[b]{\smash{{{\SetFigFont{9}{10.8}{rm}odd}}}}}
\put(2734,1901){\makebox(0,0)[b]{\smash{{{\SetFigFont{9}{10.8}{rm}even}}}}}
\put(7077,1924){\makebox(0,0)[b]{\smash{{{\SetFigFont{12}{14.4}{rm}$C'$}}}}}
\put(7032,79){\makebox(0,0)[b]{\smash{{{\SetFigFont{12}{14.4}{rm}$C$}}}}}
\put(4917,1406){\makebox(0,0)[b]{\smash{{{\SetFigFont{9}{10.8}{rm}$1$}}}}}
\put(2037,506){\makebox(0,0)[b]{\smash{{{\SetFigFont{9}{10.8}{rm}$1$}}}}}
\put(687,506){\makebox(0,0)[b]{\smash{{{\SetFigFont{9}{10.8}{rm}$1$}}}}}
\put(4737,1743){\blacken\ellipse{90}{90}}
\put(4737,1743){\ellipse{90}{90}}
\end{picture}
}
\caption{The cycle on $8$ vertices described in Lemma \ref{Lem:C8dansPermutationGraph}} \label{Fig:C8dansPermutationGraph}
\end{figure}
\begin{lem}\label{Lem:C8dansPermutationGraph}
Let $G$ be a permutation graph with a $2$-factor containing precisely two odd cycles $C$ and $C'$.
Assume that $\chi'(G)=4$ and that $(C,C')$ is not a good pair.
Let $ab$ be an edge of $C$ such that the odd path determined on $C'$ with the neighbors of $a$
and $b$, say $a'$ and $b'$ respectively, has minimum length. Assume that $C$ and $C'$ have an
orientation such that $C(a,b)$ is an edge and $C'(a',b')$ has odd length.\\
Then there must exist $4$ additional vertices $c$ and $d$ on $C$ and their neighbors on $C'$, say $c'$ and $d'$ respectively, verifying~:
\begin{itemize}
\item the paths $C'(a',d')$, $C'(b',c')$ and $C(d,c)$ are edges.
\item the path $C(b,d)$ is odd and the path $C'(d',b')$ is even.
\end{itemize}
\end{lem}
\begin{prf}
Observe first that $a'$ and $b'$ are not adjacent otherwise the cycle obtained with the paths
$C(b,a)$ and $C'(b',a')$ together with the edges $aa'$ and $bb'$ would be hamiltonian, a contradiction since it is assumed that $\chi'(G)=4$.

Since the path $C'(a',b')$ is odd there must be a neighbor of $b'$ on $C'(b',a')$, say $c'$.
Let $c$ be the neighbor of $c'$ on $C$. The path $C(b,c)$ has even length, otherwise $(aa',bb',cc')$ would
be a good triple and $(C,C')$ a good pair, a contradiction.

It follows that the vertex $c$ has a neighbor, say $d$ on $C(b,c)$ and $C(b,d)$ has odd length.

Let $d'$ be the neighbor of $d$ on $C'$. It must be pointed out that $d'$ is a vertex of $C'(a',b')$.
As a matter of fact if on the contrary $d'$ belongs to $C'(c',a')$ we would have a good triple with
$(dd', cc', bb')$ when $C'(c',d')$ has odd length and with $(aa', bb', dd')$ when $C'(c', d')$ is an even path; a contradiction in both cases.

But now by the choice of the edge $ab$ the length of $C'(a',b')$ cannot be greater than $C'(d',c')$, thus
$d'$ is adjacent to $a'$ and the path $C'(d',b')$ has even length.
\end{prf}

We have :
\begin{thm}\label{Theorem:PermutationTau4} Let $G$ be a permutation
graph then $\tau(G) \leq 4$ or $G$ is the Petersen graph.
\end{thm}
\begin{prf}
Let $C$ and $C'$ the $2-$factor of chordless cycles which partition $V(G)$ and
We can assume that $G$ is not $3-$edge colourable otherwise $\tau(G)=3$ and there is nothing to prove. Hence, $C$ and $C'$ have both odd
lengths. In addition we assume that $(C,C')$ is not a good pair, otherwise we are done by Theorem \ref{Theorem:2Factor_GoodPair}.

\medskip Let $x_1x_2$ be an edge of $C$ such that the odd path determined on $C'$ with the neighbors of
$x_1$ and $x_2$, say $y_1$ and $y_2$ respectively, has minimum length.

We choose to orient $C$ from $x_1$ to $x_2$ and to orient $C'$  from $y_1$ to $y_2$. Thus $C(x_1,x_2)$
is an edge and $C'(y_1,y_2)$ is an odd path.

\medskip By Lemma \ref{Lem:C8dansPermutationGraph} we must have two vertices $x_3$ and $x_4$ on $C$ and
their neighbors $y_3$ and $y_4$ on $C'$ such that $C(x_4, x_3)$, $C'(y_1, y_4)$, $C'(y_2, y'_3)$ are edges,
$C(x_2, x_4)$ being an odd path while $C'(y_4, y_2)$ has even length.

\begin{Clm}\label{Lemma:FondamentalOddCycle}
The vertices $y_1$ and $y_3$ are adjacent.
\end{Clm}
\begin{PrfClaim}
Assume not.

\medskip The odd path $C'(y_4, y_3)$ having the same length than $C'(y_1, y_2)$ we may apply
Lemma \ref{Lem:C8dansPermutationGraph} on the edge $x_4x_3$ ($x_4=a$, $x_3=b$). Thus there is edges;
say $x_5y_5$ and $x_6y_6$, $x_5$ and $x_6$ being vertices of $C$, $y_5$ and $y_6$ vertices of $C'$,
the paths $C(x_6,x_5)$, $C'(y_4, y_6)$ and $C'(y_3, y_5)$ having length $1$. Moreover the paths $C(x_3, x_6)$
and $C'(y_6, y_2)$ are odd. Since it is assumed that $y_1$ and $y_3$ are independent we have $y_5\neq y_1$ and $x_5\neq x_1$.

\medskip Observe that the paths $C'(y_1, y_2)$ and $C'(y_6, y_5)$ have the same length, thus we apply
Lemma \ref{Lem:C8dansPermutationGraph} again with $a=x_6$ and $b=x_5$.

Let $y_7$ be the neighbor of $y_5$ on $C'y_5, y_1)$ and $x_7$ be the neighbor of $y_7$ on $C$. We know
that $x_7$ is a vertex of $C(x_5, x_1)$ at even distance of $x_5$. The vertex $x_8$ being the neighbor
of $x_7$ on $C(x_5, x_7)$ and $y_8$ the neighbor of $x_8$ on $C'$, we have that $y_8$ is the neighbor of $y_6$ on $C'(y_6, y_2)$.

\medskip The path $C'(y_8, y_2)$ has even length, hence there must be on this path a neighbor of $y_8$
distinct from $y_2$, say $y_9$. Let $x_9$ be the neighbor of $y_9$ on $C$.

\medskip The vertex $x_9$ belongs to $C(x_7,x_1)$.
Otherwise when  $x_9$ is on $C(x_2, x_4)$; if the path $C(x_2, x_9)$ is odd we can find a good triple, namely $(x_8y_8, x_9y_9, x_2y_2)$  on the other case we have the good triple $(x_9y_9, x_4y_4, x_1y_1)$. A contradiction in both cases.

We get a similar contradiction if $x_9$ belongs to $C(x_3, x_6)$ by considering the triples $(x_5y_5, x_9y_9, x_8y_8)$ or $(x_9y_9, x_4y_4, x_2y_2)$.

Finally, when $x_9$ is a vertex of $C(x_5, x_8)$ a contradiction occurs with the triple $(x_5y_5, x_9y_9, x_7, y_7)$ if $C(x_5, x_9)$ is odd and with the tripe $(x_8y_8, x_9y_9, x_6y_6)$ otherwise.

\medskip Observe that the path $C(x_7, x_9)$ must be odd or $(x_9y_9, x_7y_7, x_8y_8)$ would be a good triple, a contradiction.

But now $(x_9y_9, x_5y_5, x_4y_4)$ is a good triple, a contradiction which proves the Claim (see Figure \ref{Fig:ThmPermutationGraphFinClaim1}).
\begin{figure}
\setlength{\unitlength}{0.00065617in}
\begingroup\makeatletter\ifx\SetFigFont\undefined
% extract first six characters in \fmtname
\def\x#1#2#3#4#5#6#7\relax{\def\x{#1#2#3#4#5#6}}%
\expandafter\x\fmtname xxxxxx\relax \def\y{splain}%
\ifx\x\y   % LaTeX or SliTeX?
\gdef\SetFigFont#1#2#3{%
  \ifnum #1<17\tiny\else \ifnum #1<20\small\else
  \ifnum #1<24\normalsize\else \ifnum #1<29\large\else
  \ifnum #1<34\Large\else \ifnum #1<41\LARGE\else
     \huge\fi\fi\fi\fi\fi\fi
  \csname #3\endcsname}%
\else
\gdef\SetFigFont#1#2#3{\begingroup
  \count@#1\relax \ifnum 25<\count@\count@25\fi
  \def\x{\endgroup\@setsize\SetFigFont{#2pt}}%
  \expandafter\x
    \csname \romannumeral\the\count@ pt\expandafter\endcsname
    \csname @\romannumeral\the\count@ pt\endcsname
  \csname #3\endcsname}%
\fi
\fi\endgroup
{\renewcommand{\dashlinestretch}{30}
\begin{picture}(7674,2264)(0,-10)
\put(462,2036){\makebox(0,0)[b]{\smash{{{\SetFigFont{9}{10.8}{rm}$y_1$}}}}}
\put(5412,461){\blacken\ellipse{90}{90}}
\put(5412,461){\ellipse{90}{90}}
\put(4962,461){\blacken\ellipse{90}{90}}
\put(4962,461){\ellipse{90}{90}}
\put(4062,461){\blacken\ellipse{90}{90}}
\put(4062,461){\ellipse{90}{90}}
\put(3612,461){\blacken\ellipse{90}{90}}
\put(3612,461){\ellipse{90}{90}}
\put(2262,1811){\blacken\ellipse{90}{90}}
\put(2262,1811){\ellipse{90}{90}}
\put(1812,1811){\blacken\ellipse{90}{90}}
\put(1812,1811){\ellipse{90}{90}}
\put(1362,1811){\blacken\ellipse{90}{90}}
\put(1362,1811){\ellipse{90}{90}}
\put(6087,1811){\blacken\ellipse{90}{90}}
\put(6087,1811){\ellipse{90}{90}}
\put(5637,1811){\blacken\ellipse{90}{90}}
\put(5637,1811){\ellipse{90}{90}}
\put(4737,1811){\blacken\ellipse{90}{90}}
\put(4737,1811){\ellipse{90}{90}}
\put(912,472){\blacken\ellipse{90}{90}}
\put(912,472){\ellipse{90}{90}}
\put(2262,461){\blacken\ellipse{90}{90}}
\put(2262,461){\ellipse{90}{90}}
\put(1857,461){\blacken\ellipse{90}{90}}
\put(1857,461){\ellipse{90}{90}}
\put(5142,1811){\blacken\ellipse{90}{90}}
\put(5142,1811){\ellipse{90}{90}}
\put(912,1811){\blacken\ellipse{90}{90}}
\put(912,1811){\ellipse{90}{90}}
\put(462,1811){\blacken\ellipse{90}{90}}
\put(462,1811){\ellipse{90}{90}}
\put(462,461){\blacken\ellipse{90}{90}}
\put(462,461){\ellipse{90}{90}}
\thicklines
\path(5637,1811)(6087,1811)
\path(5187,1811)(5637,1811)
\path(4737,1811)(5187,1811)
\path(4962,461)(5412,461)
\path(3612,461)(4062,461)
\path(1857,461)(2307,461)
\path(507,461)(957,461)
\path(1857,1811)(2307,1811)
\path(1407,1811)(1857,1811)
\path(957,1811)(1407,1811)
\path(462,1811)(912,1811)
\thinlines
\path(2262,1811)(6492,461)
\path(1812,1811)(4962,461)
\path(1362,1811)(3612,461)
\path(6087,1811)(5412,461)
\path(5637,1811)(4062,461)
\path(4737,1811)(912,461)
\path(4737,1811)(912,461)
\path(2262,461)(5142,1811)
\path(2262,461)(5142,1811)
\path(912,1811)(1857,461)
\path(912,1811)(1857,461)
\path(462,1811)(462,461)
\path(462,1811)(462,461)
\path(12,461)(7662,461)
\blacken\path(7422.000,389.000)(7662.000,461.000)(7422.000,533.000)(7494.000,461.000)(7422.000,389.000)
\path(12,1811)(7662,1811)
\blacken\path(7422.000,1739.000)(7662.000,1811.000)(7422.000,1883.000)(7494.000,1811.000)(7422.000,1739.000)
\put(5952,146){\makebox(0,0)[b]{\smash{{{\SetFigFont{9}{10.8}{rm}odd}}}}}
\put(6492,192){\makebox(0,0)[b]{\smash{{{\SetFigFont{9}{10.8}{rm}$x_9$}}}}}
\put(4962,192){\makebox(0,0)[b]{\smash{{{\SetFigFont{9}{10.8}{rm}$x_8$}}}}}
\put(5412,192){\makebox(0,0)[b]{\smash{{{\SetFigFont{9}{10.8}{rm}$x_7$}}}}}
\put(3612,192){\makebox(0,0)[b]{\smash{{{\SetFigFont{9}{10.8}{rm}$x_6$}}}}}
\put(4062,192){\makebox(0,0)[b]{\smash{{{\SetFigFont{9}{10.8}{rm}$x_5$}}}}}
\put(5187,574){\makebox(0,0)[b]{\smash{{{\SetFigFont{9}{10.8}{rm}$1$}}}}}
\put(3837,574){\makebox(0,0)[b]{\smash{{{\SetFigFont{9}{10.8}{rm}$1$}}}}}
\put(4512,146){\makebox(0,0)[b]{\smash{{{\SetFigFont{9}{10.8}{rm}odd}}}}}
\put(2892,146){\makebox(0,0)[b]{\smash{{{\SetFigFont{9}{10.8}{rm}odd}}}}}
\put(2239,2036){\makebox(0,0)[b]{\smash{{{\SetFigFont{9}{10.8}{rm}$y_9$}}}}}
\put(1789,2036){\makebox(0,0)[b]{\smash{{{\SetFigFont{9}{10.8}{rm}$y_8$}}}}}
\put(1384,2036){\makebox(0,0)[b]{\smash{{{\SetFigFont{9}{10.8}{rm}$y_6$}}}}}
\put(3274,1924){\makebox(0,0)[b]{\smash{{{\SetFigFont{9}{10.8}{rm}odd}}}}}
\put(2037,1564){\makebox(0,0)[b]{\smash{{{\SetFigFont{9}{10.8}{rm}$1$}}}}}
\put(1632,1564){\makebox(0,0)[b]{\smash{{{\SetFigFont{9}{10.8}{rm}$1$}}}}}
\put(1182,1564){\makebox(0,0)[b]{\smash{{{\SetFigFont{9}{10.8}{rm}$1$}}}}}
\put(6064,2036){\makebox(0,0)[b]{\smash{{{\SetFigFont{9}{10.8}{rm}$y_7$}}}}}
\put(5659,2036){\makebox(0,0)[b]{\smash{{{\SetFigFont{9}{10.8}{rm}$y_5$}}}}}
\put(5412,1474){\makebox(0,0)[b]{\smash{{{\SetFigFont{9}{10.8}{rm}$1$}}}}}
\put(5862,1474){\makebox(0,0)[b]{\smash{{{\SetFigFont{9}{10.8}{rm}$1$}}}}}
\put(1317,191){\makebox(0,0)[b]{\smash{{{\SetFigFont{9}{10.8}{rm}odd}}}}}
\put(7527,2014){\makebox(0,0)[b]{\smash{{{\SetFigFont{12}{14.4}{rm}$C'$}}}}}
\put(7527,79){\makebox(0,0)[b]{\smash{{{\SetFigFont{12}{14.4}{rm}$C$}}}}}
\put(889,2036){\makebox(0,0)[b]{\smash{{{\SetFigFont{9}{10.8}{rm}$y_4$}}}}}
\put(5164,2036){\makebox(0,0)[b]{\smash{{{\SetFigFont{9}{10.8}{rm}$y_3$}}}}}
\put(4737,2036){\makebox(0,0)[b]{\smash{{{\SetFigFont{9}{10.8}{rm}$y_2$}}}}}
\put(2307,192){\makebox(0,0)[b]{\smash{{{\SetFigFont{9}{10.8}{rm}$x_3$}}}}}
\put(1879,214){\makebox(0,0)[b]{\smash{{{\SetFigFont{9}{10.8}{rm}$x_4$}}}}}
\put(4917,1474){\makebox(0,0)[b]{\smash{{{\SetFigFont{9}{10.8}{rm}$1$}}}}}
\put(2037,574){\makebox(0,0)[b]{\smash{{{\SetFigFont{9}{10.8}{rm}$1$}}}}}
\put(687,574){\makebox(0,0)[b]{\smash{{{\SetFigFont{9}{10.8}{rm}$1$}}}}}
\put(687,1564){\makebox(0,0)[b]{\smash{{{\SetFigFont{9}{10.8}{rm}$1$}}}}}
\put(889,192){\makebox(0,0)[b]{\smash{{{\SetFigFont{9}{10.8}{rm}$x_2$}}}}}
\put(462,192){\makebox(0,0)[b]{\smash{{{\SetFigFont{9}{10.8}{rm}$x_1$}}}}}
\put(6492,461){\blacken\ellipse{90}{90}}
\put(6492,461){\ellipse{90}{90}}
\end{picture}
}
\caption{Situation at the end of Claim \ref{Theorem:PermutationTau4:clm2}} \label{Fig:ThmPermutationGraphFinClaim1}
\end{figure}
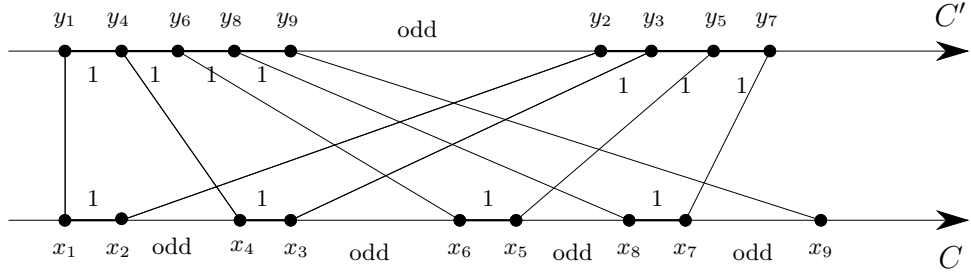
\end{PrfClaim}

From now on we assume that $y_3y_1$ is an edge.

\medskip The path $C(x_1, x_3)$ being odd there must be a neighbor of $x_3$ on $C(x_3,x_1)$ distinct from
$x_1$, let $x_5$ be this vertex. It's neighbor on $C'$, say $y_5$, must be on $C'(y_4, y_2)$. Moreover the
length of $C'(y_4, y_5)$ is odd otherwise the edges $x_5y_5$, $x_3y_3$ and $x_1y_1$ would form a good triple, a contradiction.
\begin{Clm}\label{Theorem:PermutationTau4:clm2}
The paths $C'(y_4, y_5)$ and $C'(y_5, y_2)$ are reduced to edges.
\end{Clm}
\begin{PrfClaim}
Assume in a first stage that the neighbor of $y_4$ on $C'(y_4, y_5)$ is distinct from $y_5$, let $y_6$ be
this vertex and $x_6$ be its neighbor on $C$.

The vertex $x_6$ cannot belong to $C(x_5, x_1)$, otherwise we would have a good triple $(x_3y_3, x_6y_6, x_4y_4)$
when $C(x_5, x_6)$ is an even path and the good triple $(x_4y_4, x_6y_6, x_2y_2)$ if it's an odd path, contradictions.

Similarly the vertex $x_6$ cannot belong to $C(x_2, x_4)$. On the contrary we would have a good triple with the
edges $x_2y_2$, $x_6y_6$ and $x_1y_1$ when the path $C(x_2, x_6)$ is odd and another good triple with the edges $x_4y_4$, $x_6y_6$ and $x_1y_1$.

\medskip  On the same manner we can prove that the path $C'(y_5, y_2)$ has length $1$.
\end{PrfClaim}

It comes from Claim \ref{Theorem:PermutationTau4:clm2} that $C'$ has only $5$ vertices. Since both cycles $C$
and $C'$ have the same length $C$ has $5$ vertices too and $G$ is the Petersen graph.
\end{prf}
\begin{figure}
\begin{center}
\setlength{\unitlength}{0.00052493in}
\begingroup\makeatletter\ifx\SetFigFontNFSS\undefined%
\gdef\SetFigFontNFSS#1#2#3#4#5{%
  \reset@font\fontsize{#1}{#2pt}%
  \fontfamily{#3}\fontseries{#4}\fontshape{#5}%
  \selectfont}%
\fi\endgroup%
{\renewcommand{\dashlinestretch}{30}
\begin{picture}(11004,2142)(0,-10)
\put(1362,1743){\blacken\ellipse{128}{128}}
\put(1362,1743){\ellipse{128}{128}}
\put(687,1743){\blacken\ellipse{128}{128}}
\put(687,1743){\ellipse{128}{128}}
\put(2037,1743){\blacken\ellipse{128}{128}}
\put(2037,1743){\ellipse{128}{128}}
\put(687,1068){\blacken\ellipse{128}{128}}
\put(687,1068){\ellipse{128}{128}}
\put(687,393){\blacken\ellipse{128}{128}}
\put(687,393){\ellipse{128}{128}}
\put(2037,1068){\blacken\ellipse{128}{128}}
\put(2037,1068){\ellipse{128}{128}}
\put(2037,393){\blacken\ellipse{128}{128}}
\put(2037,393){\ellipse{128}{128}}
\put(1362,393){\blacken\ellipse{128}{128}}
\put(1362,393){\ellipse{128}{128}}
\path(687,1068)(2037,1068)
\path(1362,1743)(1362,393)
\path(687,1743)(2037,1743)(2037,393)
	(687,393)(687,1743)
\path(687,1743)(12,1743)
\path(687,393)(12,393)
\path(2037,1743)(2712,1743)
\path(2037,393)(2712,393)
\put(5187,1743){\blacken\ellipse{128}{128}}
\put(5187,1743){\ellipse{128}{128}}
\put(4512,1743){\blacken\ellipse{128}{128}}
\put(4512,1743){\ellipse{128}{128}}
\put(5862,1743){\blacken\ellipse{128}{128}}
\put(5862,1743){\ellipse{128}{128}}
\put(4512,1068){\blacken\ellipse{128}{128}}
\put(4512,1068){\ellipse{128}{128}}
\put(4512,393){\blacken\ellipse{128}{128}}
\put(4512,393){\ellipse{128}{128}}
\put(5862,1068){\blacken\ellipse{128}{128}}
\put(5862,1068){\ellipse{128}{128}}
\put(5862,393){\blacken\ellipse{128}{128}}
\put(5862,393){\ellipse{128}{128}}
\put(5187,393){\blacken\ellipse{128}{128}}
\put(5187,393){\ellipse{128}{128}}
\put(6537,1743){\blacken\ellipse{128}{128}}
\put(6537,1743){\ellipse{128}{128}}
\put(6537,393){\blacken\ellipse{128}{128}}
\put(6537,393){\ellipse{128}{128}}
\put(9642,1743){\blacken\ellipse{128}{128}}
\put(9642,1743){\ellipse{128}{128}}
\put(8967,1743){\blacken\ellipse{128}{128}}
\put(8967,1743){\ellipse{128}{128}}
\put(10317,1743){\blacken\ellipse{128}{128}}
\put(10317,1743){\ellipse{128}{128}}
\put(8967,1068){\blacken\ellipse{128}{128}}
\put(8967,1068){\ellipse{128}{128}}
\put(8967,393){\blacken\ellipse{128}{128}}
\put(8967,393){\ellipse{128}{128}}
\put(10317,1068){\blacken\ellipse{128}{128}}
\put(10317,1068){\ellipse{128}{128}}
\put(10317,393){\blacken\ellipse{128}{128}}
\put(10317,393){\ellipse{128}{128}}
\put(9642,393){\blacken\ellipse{128}{128}}
\put(9642,393){\ellipse{128}{128}}
\put(9867,1068){\blacken\ellipse{128}{128}}
\put(9867,1068){\ellipse{128}{128}}
\put(9642,1293){\blacken\ellipse{128}{128}}
\put(9642,1293){\ellipse{128}{128}}
\path(4512,1068)(5862,1068)
\path(5187,1743)(5187,393)
\path(4512,1743)(5862,1743)(5862,393)
	(4512,393)(4512,1743)
\path(5862,1743)(6537,1743)
\path(5862,393)(7212,393)
\path(6537,1743)(7212,1743)
\path(6537,1743)(6537,393)
\path(4512,1743)(3837,1743)
\path(4512,393)(3837,393)
\path(8967,1068)(10317,1068)
\path(9642,1743)(9642,393)
\path(8967,1743)(10317,1743)(10317,393)
	(8967,393)(8967,1743)
\path(8967,1743)(8292,1743)
\path(8967,393)(8292,393)
\path(10317,1743)(10992,1743)
\path(10317,393)(10992,393)
\path(9642,1293)(9867,1068)
\put(102,1968){\makebox(0,0)[lb]{\smash{{\SetFigFontNFSS{7}{8.4}{\rmdefault}{\mddefault}{\updefault}$a$}}}}
\put(2262,1923){\makebox(0,0)[lb]{\smash{{\SetFigFontNFSS{7}{8.4}{\rmdefault}{\mddefault}{\updefault}$b'$}}}}
\put(2352,483){\makebox(0,0)[lb]{\smash{{\SetFigFontNFSS{7}{8.4}{\rmdefault}{\mddefault}{\updefault}$a'$}}}}
\put(192,483){\makebox(0,0)[lb]{\smash{{\SetFigFontNFSS{7}{8.4}{\rmdefault}{\mddefault}{\updefault}$b$}}}}
\put(8382,1968){\makebox(0,0)[lb]{\smash{{\SetFigFontNFSS{7}{8.4}{\rmdefault}{\mddefault}{\updefault}$a$}}}}
\put(10542,1923){\makebox(0,0)[lb]{\smash{{\SetFigFontNFSS{7}{8.4}{\rmdefault}{\mddefault}{\updefault}$b'$}}}}
\put(10632,483){\makebox(0,0)[lb]{\smash{{\SetFigFontNFSS{7}{8.4}{\rmdefault}{\mddefault}{\updefault}$a'$}}}}
\put(8472,483){\makebox(0,0)[lb]{\smash{{\SetFigFontNFSS{7}{8.4}{\rmdefault}{\mddefault}{\updefault}$b$}}}}
\put(9642,78){\makebox(0,0)[b]{\smash{{\SetFigFontNFSS{7}{8.4}{\rmdefault}{\mddefault}{\updefault}Block $A_2$}}}}
\put(5322,123){\makebox(0,0)[b]{\smash{{\SetFigFontNFSS{7}{8.4}{\rmdefault}{\mddefault}{\updefault}Block $A_1$}}}}
\put(1407,123){\makebox(0,0)[b]{\smash{{\SetFigFontNFSS{7}{8.4}{\rmdefault}{\mddefault}{\updefault}Block $B$}}}}
\put(4017,483){\makebox(0,0)[lb]{\smash{{\SetFigFontNFSS{7}{8.4}{\rmdefault}{\mddefault}{\updefault}$b$}}}}
\put(3972,1923){\makebox(0,0)[lb]{\smash{{\SetFigFontNFSS{7}{8.4}{\rmdefault}{\mddefault}{\updefault}$a$}}}}
\put(6852,483){\makebox(0,0)[lb]{\smash{{\SetFigFontNFSS{7}{8.4}{\rmdefault}{\mddefault}{\updefault}$a'$}}}}
\put(6807,1923){\makebox(0,0)[lb]{\smash{{\SetFigFontNFSS{7}{8.4}{\rmdefault}{\mddefault}{\updefault}$b'$}}}}
\end{picture}
}
\caption{Blocks for the construction of generalized Blanu\v{s}a snarks.}
\label{Fig:Blanusa-Blocks}
\end{center}
\end{figure}
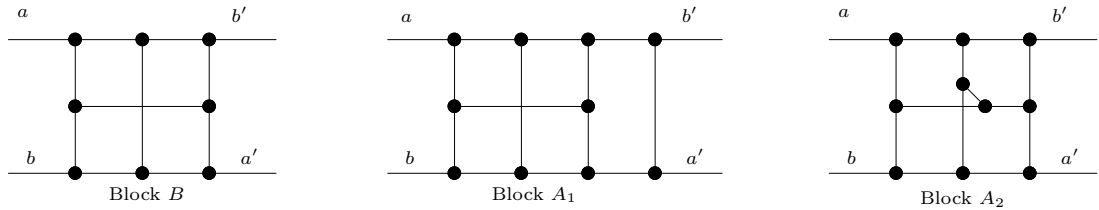

In \cite{Wat89} Watkins proposed two families of generalized Blanu\v{s}a snarks using the blocks $B$, $A_1$ and $A_2$ described in Figure \ref{Fig:Blanusa-Blocks}. The generalized Blanu\v{s}a snarks of type $1$ (resp. of type $2$) are obtained by considering a number of blocks $B$ and one block $A_1$ (resp. $A_2$), these blocks are arranged cyclically, the semi-edges $a$ and  $b$ of one block being connected to the semi-edges $a$, $b$ of the next one. Recently generalized Blanu\v{s}a snarks were studied in terms of circular chromatic index (see \cite{Maz08,Ghe08}).

The generalized Blanu\v{s}a snarks are permutation graphs, hence~:
\begin{cor}\label{cor:generalized Blanusa snarks}
Let $G$ be a generalized Blanu\v{s}a snarks then $\tau(G)=4$.
\end{cor}

\section{\label{Section:Tau5} On graphs
with $\tau \geq 5$}
It is an easy task to construct cubic graphs
with perfect matching index at least $5$ with the help of
Proposition \ref{Proposition:Construction:2Cut}. Take indeed the
Petersen graph $P$ and any bridgeless cubic graph $G$ and apply the
construction $P\bigodot G$.

\begin{prop}\label{Proposition:Constructiontau5withbiparti} Let $G$ be  bridgeless cubic graph
with perfect matching index at least  $5$ and let $H$ be a
connected bipartite cubic graph. Then $G \otimes H$ is bridgeless
cubic graph with perfect matching index at least  $5$.
\end{prop}
\begin{prf}
Assume that $\tau(G \otimes H) = 4$ and let $\mathcal M=\{M_{1},
M_{2},M_{3},M_{4}\}$ be a covering of its edge set into $4$ perfect
matchings. Let $\{aa',bb',cc'\}$ (with $a,b$ and $c$ in $G$ and
$a',b'$ and $c'$ in $H$) be the principal $3-$edge cut of $G
\otimes H$. From Item \ref{Item:Construction:3OddCutTwo} of
Proposition \ref{Proposition:Construction:3OddCut} there is perfect
matching $M_{i} \in \mathcal M$ such that $\{aa',bb',cc'\} \subseteq
M_{i}$. This is clearly impossible since the set of vertices of $H$
which must be saturated by $M_{i}$ is partitioned into $2$
independent sets whose size differs by one unit.
\end{prf}

\begin{figure}
\begin{center}
\setlength{\unitlength}{0.00065617in}
\begingroup\makeatletter\ifx\SetFigFont\undefined
% extract first six characters in \fmtname
\def\x#1#2#3#4#5#6#7\relax{\def\x{#1#2#3#4#5#6}}%
\expandafter\x\fmtname xxxxxx\relax \def\y{splain}%
\ifx\x\y   % LaTeX or SliTeX?
\gdef\SetFigFont#1#2#3{%
  \ifnum #1<17\tiny\else \ifnum #1<20\small\else
  \ifnum #1<24\normalsize\else \ifnum #1<29\large\else
  \ifnum #1<34\Large\else \ifnum #1<41\LARGE\else
     \huge\fi\fi\fi\fi\fi\fi
  \csname #3\endcsname}%
\else
\gdef\SetFigFont#1#2#3{\begingroup
  \count@#1\relax \ifnum 25<\count@\count@25\fi
  \def\x{\endgroup\@setsize\SetFigFont{#2pt}}%
  \expandafter\x
    \csname \romannumeral\the\count@ pt\expandafter\endcsname
    \csname @\romannumeral\the\count@ pt\endcsname
  \csname #3\endcsname}%
\fi
\fi\endgroup
{\renewcommand{\dashlinestretch}{30}
\begin{picture}(5315,4475)(0,-10)
\put(565,610){\makebox(0,0)[lb]{\smash{{{\SetFigFont{9}{10.8}{rm}$G_1$}}}}}
\put(3130,1510){\blacken\ellipse{66}{66}}
\put(3130,1510){\ellipse{66}{66}}
\put(1240,1150){\blacken\ellipse{66}{66}}
\put(1240,1150){\ellipse{66}{66}}
\put(2590,2365){\blacken\ellipse{66}{66}}
\put(2590,2365){\ellipse{66}{66}}
\put(2230,1555){\blacken\ellipse{66}{66}}
\put(2230,1555){\ellipse{66}{66}}
\put(2635,1825){\ellipse{1654}{1654}}
\put(2266,3252){\blacken\ellipse{66}{66}}
\put(2266,3252){\ellipse{66}{66}}
\put(2581,3072){\blacken\ellipse{66}{66}}
\put(2581,3072){\ellipse{66}{66}}
\put(3007,3220){\blacken\ellipse{66}{66}}
\put(3007,3220){\ellipse{66}{66}}
\put(4582,1465){\blacken\ellipse{66}{66}}
\put(4582,1465){\ellipse{66}{66}}
\put(4132,835){\blacken\ellipse{66}{66}}
\put(4132,835){\ellipse{66}{66}}
\put(1252,835){\blacken\ellipse{66}{66}}
\put(1252,835){\ellipse{66}{66}}
\put(736,1362){\blacken\ellipse{66}{66}}
\put(736,1362){\ellipse{66}{66}}
\put(4480,835){\ellipse{1654}{1654}}
\put(835,835){\ellipse{1654}{1654}}
\put(2545,3625){\ellipse{1654}{1654}}
\path(3130,1510)(4075,1150)
\path(3130,1510)(4075,1150)
\path(2590,3040)(2590,2365)
\path(2590,3040)(2590,2365)
\path(1240,1150)(2230,1555)
\path(1240,1150)(2230,1555)
\path(1285,835)(4120,835)
\path(1285,835)(4120,835)
\path(4570,1465)(2995,3220)
\path(4570,1465)(2995,3220)
\path(745,1375)(2275,3265)
\path(745,1375)(2275,3265)
\put(2365,1780){\makebox(0,0)[lb]{\smash{{{\SetFigFont{9}{10.8}{rm}$G_4$}}}}}
\put(2455,2185){\makebox(0,0)[lb]{\smash{{{\SetFigFont{7}{8.4}{rm}$b_4$}}}}}
\put(2860,1330){\makebox(0,0)[lb]{\smash{{{\SetFigFont{7}{8.4}{rm}$c_4$}}}}}
\put(2185,1330){\makebox(0,0)[lb]{\smash{{{\SetFigFont{7}{8.4}{rm}$a_4$}}}}}
\put(2905,3355){\makebox(0,0)[lb]{\smash{{{\SetFigFont{7}{8.4}{rm}$a_1$}}}}}
\put(430,1420){\makebox(0,0)[lb]{\smash{{{\SetFigFont{7}{8.4}{rm}$a_1$}}}}}
\put(3895,655){\makebox(0,0)[lb]{\smash{{{\SetFigFont{7}{8.4}{rm}$c_2$}}}}}
\put(4345,1060){\makebox(0,0)[lb]{\smash{{{\SetFigFont{7}{8.4}{rm}$b_2$}}}}}
\put(4750,1375){\makebox(0,0)[lb]{\smash{{{\SetFigFont{7}{8.4}{rm}$a_2$}}}}}
\put(2500,3175){\makebox(0,0)[lb]{\smash{{{\SetFigFont{7}{8.4}{rm}$b_3$}}}}}
\put(2005,3355){\makebox(0,0)[lb]{\smash{{{\SetFigFont{7}{8.4}{rm}$c_3$}}}}}
\put(1240,655){\makebox(0,0)[lb]{\smash{{{\SetFigFont{7}{8.4}{rm}$c_1$}}}}}
\put(745,1060){\makebox(0,0)[lb]{\smash{{{\SetFigFont{7}{8.4}{rm}$b_1$}}}}}
\put(2320,3670){\makebox(0,0)[lb]{\smash{{{\SetFigFont{9}{10.8}{rm}$G_3$}}}}}
\put(4300,565){\makebox(0,0)[lb]{\smash{{{\SetFigFont{9}{10.8}{rm}$G_2$}}}}}
\put(4087,1150){\blacken\ellipse{66}{66}}
\put(4087,1150){\ellipse{66}{66}}
\end{picture}
}
\end{center}
\caption{$K_4[G_1^{x_1}, G_2^{x_2}, G_3^{x_3}, G_4^{x_4}]$}\label{Fig:K4Composition}
\end{figure}
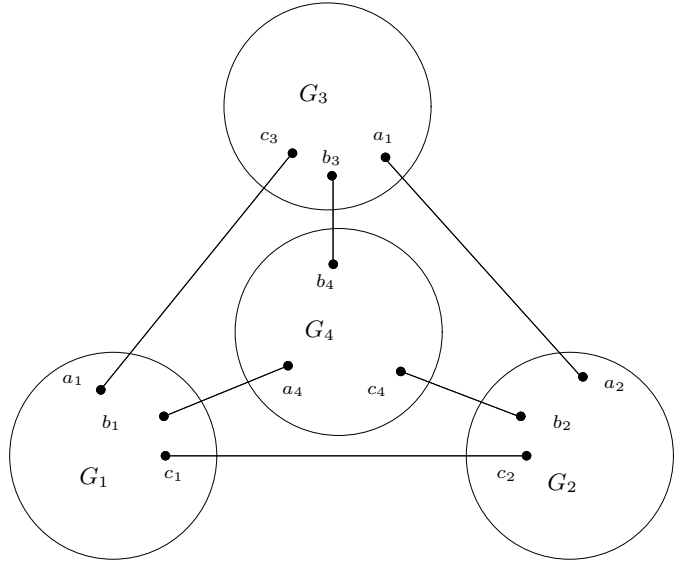

Let us consider the following construction. Given four cubic graphs $G_1^{x_1}$, $G_2^{x_2}$, $G_3^{x_3}$, $G_4^{x_4}$ together with a distinguished vertex $x_i$ ($i=1,2,3,4$) whose neighbors in $G_i^{x_i}$ are $a_i$, $b_i$ and $c_i$, we get a $3$-connected cubic graphs in deleting the vertices $x_i$ ($i=1,2,3,4$) and connecting the remaining subgraphs as described in Figure \ref{Fig:K4Composition}. In other words we define the cubic graphs denoted $K_4[G_1^{x_1}, G_2{x_2}, G_3^{x_3}, G_4^{x_4}]$ whose vertex set is $$\bigcup_{i\in\{1,2,3,4\}}V(G_i^{x_i})-\bigcup_{i\in\{1,2,3,4\}} \{x_i\}$$ while the edge set is $$\bigcup_{i\in\{1,2,3,4\}}E(G_i^{x_i})-\bigcup_{i\in\{1,2,3,4\}}\{a_ix_i,b_ix_i,c_ix_i\}\bigcup \{a_1c_3, b_1a_4, c_1c_2, b_2c_4, a_2c_3,b_3b_4\}.$$
For convenience $G_i$ ($i\in\{1,2,3,4\}$) will denote the induced subgraph of $G_i^{x_i}$ where the vertex $x_i$ has been deleted.

\begin{prop}\label{Proposition:Constructiontau5}
Let $G_1^{x_1}$, $G_2^{x_2}$, $G_3^{x_3}$ and $G_4^{x_4}$ be $3$-connected cubic graphs such that $\tau(G_1^{x_1})\geq 5$, $\tau(G_2^{x_2})\geq 5$, $G_4$ is reduced to a single vertex, say $x$. Then $\tau(K_4[G_1^{x_1},G_2^{x_2}, G_3^{x_3}, G_4^{y}])\geq 5$.
\end{prop}
\begin{prf}
Let us denote  $G=K_4[G_1^{x_1},G_2^{x_2}, G_3^{x_3}, G_4^{x_4}]$. Observe that $a_4=b_4=c_4=x$.

If $\tau(G)=3$ the graph $G$ would be $3$-edge colourable, but in considering the $3$-edge cut $\{a_1a_3, b_1a_4, c_1c_2\}$ we would have $\chi'(G_1^{x_1})=3$, a contradiction. Hence $\tau(G) \geq 4$. Assume that $\tau(G) = 4$ and let $\mathcal M=\{M_{1}, M_{2},M_{3},M_{4}\}$ be a covering of its edge set into $4$ perfect matchings.

From Item \ref{Item:Construction:3OddCutTwo} of Proposition
\ref{Proposition:Construction:3OddCut} there is perfect matching
$M_{i} \in \mathcal M$ such that $\{a_1a_3, b_1a_4, c_1c_2\} \subseteq M_{i}$. For
the same reason, there is perfect matching $M_{j} \in \mathcal M$
such that $\{c_1c_2, xb_2c_3a_2\} \subseteq M_{j}$. We certainly have $i \neq j$,
otherwise the vertex $x$ is incident twice to the same perfect
matching $M_{i}$. Without loss of generality, we suppose that $i=1$
and $j=2$. Hence $c_1c_2 \in M_{1} \cap M_{2}$. If we consider the
$3-$edge cut $\{a_1a_3, b_1a_4, c_1c_2\}$, since each perfect matching must intersect this cut in an odd number of edges we must have one of the edges $a_1a_3$
or $b_1x$ in $M_{3}$ while the other must be in $M_{4}$. The same holds
with the $3-$edge cut $\{c_1c_2, xb_2c_3a_2\}$ and the edges $b_2x$ and $a_2c_3$. Hence,
we can suppose that $a_1a_3 \in M_{1} \cap M_{3}$ and $b_1x \in M_{1} \cap
M_{4}$ as well that $b_2x \in M_{2} \cap M_{3}$ and $a_2c_3 \in M_{2} \cap
M_{4}$, a contradiction since  the set of edges contained into $2$
perfect matchings of $\mathcal M$ is a perfect matching by Item
\ref{Item:TwoPerfectMatching} of Proposition
\ref{Proposition:Tau4Elementaire} and  $x$ is incident to two such
edges.
\end{prf}

%\begin{figure}[htb]
%\centering \epsfsize=0.6 \hsize \noindent
%\epsfbox{compositiontau5.eps} \caption{A graph with perfect matching
%index $\geq 5$} \label{Figure:ConstructionTau5}
%\end{figure}

We do not know any cyclically $4-$edge connected cubic graph,
distinct from the Petersen graph, having a perfect matching index
at least $5$ and we propose as an open problem:

\begin{prob}\label{Problem:Noc4ctau5DistinctFromPetersen}
Is there any  cyclically $4-$edge connected cubic graph distinct
from the Petersen graph with a perfect matching index at least $5$?
\end{prob}
\section{\label{Section:TechnicalTool5} Technical tools.}

In fact Theorem \ref{Theorem:2Factor_GoodPair} can be generalized.
Let $M$ be a perfect matching, a set $A \subseteq E(G)$ is a {\em
$M-$balanced matching} when we can find a perfect matchings $M'$
 such that $A=M \cap M'$. Assume that $\mathcal M=\{A,B,C\}$  are $3$
 pairwise disjoint $M-$balanced matchings, we shall say that
 $\mathcal M$ is a {\em good family} whenever the two following
 conditions are fulfilled:

 \begin{itemize}
   \item [i \label{Def:GoodFamily:Item:1}] Every odd cycle $C$ of $G \backslash M$ has exactly one vertex
   incident with one edge of each subset of $\mathcal M$ and the
   three paths determined by these vertices on $C$ are odd.
   \item [ii \label{Def:GoodFamily:Item:2}] For every even cycle of $G \backslash M$ there are at least two matchings of $\mathcal M$ with no edge incident to the cycle.

 \end{itemize}

\begin{thm} \label{Theorem:TechnicalTool1}  Let $G$ be a bridgeless cubic graph together with a
good family $\mathcal M$. Then $\tau(G) \leq 4$.
\end{thm}
\begin{prfsketch}
Let us denote $M_A$ (resp. $M_B$, $M_C$) a perfect matching such that $M_A\cap M=A$ (resp. $M_B\cap M=B$ , $M_C\cap M=C$).

Let $\mathcal C$ be a cycle of the $2$-factor $G-M$.

When $\mathcal C$ is an even cycle, there are precisely two matchings on $\mathcal C$, namely $M_{\mathcal C}$ and $M'_{\mathcal C}$ such that $M_{\mathcal C}\cup M'_{\mathcal C}$ covers all the edge-set of $\mathcal C$. Since there are at least two matchings in $\{M_A, M_B, M_C\}$ that are not incident to $\mathcal C$, say $M_A$ and $M_B$, up to a redistribution of the edges in $M_A\cap {\mathcal C}$ and $M_B\cap {\mathcal C}$ we may assume that $M_{\mathcal C}\subset M_A$ and $M'_{\mathcal C}\subset M_B$.

If $\mathcal C$ is an odd cycle we know that $\mathcal C$ has precisely one vertex which is incident to $A$ say $a$,
 one vertex which is incident to $B$ say $b$, one vertex which is incident to $C$ say $c$. Without loss of generality we may assume that there is an orientation of $\mathcal C$ such that the path $\mathcal C(a,b)$ has odd length and the vertex $c$ in $\mathcal C(b,a)$. We know that the path $\mathcal C(b,c)$ is odd thus the edge-set of $\mathcal C$ is covered with $M_A\cup M_B\cup M_C$.
\end{prfsketch}

In the same manner we can obtain a theorem insuring the existence of
a $5$-covering.

Assume that $\mathcal M=\{A,B,C,D\}$  are $4$
 pairwise disjoint $M-$balanced matchings, we shall say that
 $\mathcal M$ is a {\em ice family} whenever the two following
 conditions are fulfilled:

 \begin{itemize}
   \item [i \label{Def:VeryGoodFamily:Item:1}] Every odd cycle $C$ of $G \backslash M$ has exactly one vertex
   incident with one edge of each subset of $\mathcal M$ and at least two
   disjoint
    paths determined by these vertices on $C$ are odd.
   \item [ii \label{Def:VeryGoodFamily:Item:2}] For every even cycle of $G \backslash M$ there are at least two matchings of $\mathcal M$with no edge incident to the cycle.
 \end{itemize}

\begin{thm} \label{Theorem:TechnicalTool2}  Let $G$ be a bridgeless cubic graph together with a
nice family $\mathcal M$. Then $\tau(G) \leq 5$.
\end{thm}
\begin{prf}
Let us denote $M_A$ (resp. $M_B$, $M_C$, $M_D$) a perfect matching such that $M_A\cap M=A$ (resp. $M_B\cap M=B$ ,$M_C\cap M=C$,$M_D\cap M=D$).

Let $\mathcal C$ be a cycle of the $2$-factor $G-M$.

When $\mathcal C$ is an even cycle, there is at least two matchings in $\{M_A, M_B, M_C, M_D\}$ that are not incident to $\mathcal C$, say $M_1$ and $M_2$. As in Theorem \ref{Theorem:TechnicalTool1} we may assume that the edge-set of $\mathcal C$ is a subset of $M_1\cup M_2$.

If $\mathcal C$ is an odd cycle we know that $\mathcal C$ has precisely one vertex which is incident to $A$ say $a$,
 one vertex which is incident to $B$ say $b$, one vertex which is incident to $C$ say $c$, one vertex which is incident to $D$ say $d$. Without loss of generality we may assume that there is an orientation of $\mathcal C$ such that the path $\mathcal C(a,b)$ has odd length and the vertices $c$ and $d$ are in this order in $\mathcal(b,a)$. We can suppose that the path $\mathcal(b,c)$ is even otherwise the edge-set of $\mathcal C$ would be covered with $M_A\cup M_B\cup M_C$. But now, since $\mathcal C$ is an odd cycle the path $\mathcal C(d,a)$ has odd length and the edge-set of $\mathcal C$ is a subset of $M_A\cup M_B\cup M_D$ and $(M, M_A, M_B, M_C, M_D)$ is a $5$-covering.
\end{prf}

In a forthcoming paper \cite{FouVan09} we shall give an analogous
theorem insuring the existence of a Fulkerson covering and some
applications.

\section{\label{Section:Comments5} Odd or even coverings.}

A covering of a bridgeless cubic graph being a set of perfect
matchings such that every edge is contained in at least one perfect
matching, we define  an {\em odd covering}  as a covering such that
each edge is contained in an odd number of the members of the
covering. In the same way, an {\em even covering} is a covering such
that each edge is contained in an even number (at least $2$) members
of the covering. The {\em size} of an odd (or even) covering is its number of
members. As soon as a covering is given an even covering is obtained by taking each perfect matching twice.

\begin{prop}\label{Proposition:CouvertureImpaire} Let $G$ be  bridgeless cubic
graph such that $\tau(G)=4$. Then  $G$ has an odd covering of size
 $5$.
\end{prop}
\begin{prf} Let $G$ be a cubic graph such that $\tau(G)=4$
and let $\mathcal M =\{M_{1}, M_{2},M_{3},M_{4}\}$ be a covering of
its edge set into $4$ perfect matchings.   Let $M$ be the perfect
matching formed  with the edges contained in exactly two perfect
matchings of $\mathcal M$.  Then we can check that $\{M,M_{1},
M_{2},M_{3},M_{4}\}$ cover every edge of $G$ either one time or
three times.
\end{prf}

\begin{prop}\label{Proposition:CouvertureImpairePerfectMatchingsDisjoint}
Let $G$ be a bridgeless cubic graph together with an odd covering
$\mathcal M$ of size $k$. Then either  $G$ has an odd covering of
size $k-2$ or $\forall M,M' \in \mathcal M$ we have $M \not = M'$.
\end{prop}
\begin{prf}
Assume that there are two identical perfect matchings $M$ and $M'$
in $\mathcal M$. Each edge $e$ covered by $M$ (and thus  $M'$) must
be covered by at least another perfect matching $M_{e}$ and the set
$\mathcal M - \{M,M'\}$ is still an odd covering. The result
follows.
\end{prf}

\begin{prop}\label{Proposition:PeteresenSansCouvertureImpaire} The
Petersen graph has no odd covering.
\end{prop}
\begin{prf} Let $\mathcal M$ be an odd covering of the Petersen graph with minimum size.
Then, by Proposition
\ref{Proposition:CouvertureImpairePerfectMatchingsDisjoint}
$\mathcal M$ must be a set of distinct perfect matchings. The
Petersen graph has exactly $6$ distinct perfect matchings (inducing
a Fulkerson covering, that is an even covering) and it is an easy
task to check that any subset of $5$ perfect matchings is not an odd
covering. Since $\tau(Petersen)=5$, the result follows.
\end{prf}

Seymour (\cite{Sey79}) remarked that the edge set of the Petersen graph is not expressible as a symmetric difference (mod $2$) of its perfects matchings.

\begin{prob}\label{Problem:CouvertureImpaire} Which bridgeless cubic
graph can be provided with an odd covering~?
\end{prob}

We remark that  $3-$edge-colorable cubic graphs as well as
bridgeless cubic graph with perfect matching index $4$ have an odd covering
(with size $3$ and $5$ respectively).

\begin{prop}\label{Proposition:ConstructionSansCouvertureImpaire} Let $G$ be  bridgeless cubic graph
without any odd covering and let $H$ be a connected bipartite cubic
graph. Then $G \otimes H$ has no odd covering.
\end{prop}
\begin{prf}
Assume that $G \otimes H$ can be provided with an odd covering
$\mathcal M$. Let $\{aa',bb',cc'\}$ (with $a,b$ and $c$ in $G$ and
$a',b'$ and $c'$ in $H$) be the principal $3-$edge cut of $G
\otimes H$. None of the perfect matchings of $\mathcal M$ can
contain the principal $3-$edge cut since the set of vertices of $H$
which must be saturated by such a perfect matching is partitioned
into $2$ independent sets whose size differs by one unit. Hence
every perfect matching of $M \in \mathcal M$ contains exactly one
edge in $\{aa',bb',cc'\}$ and leads to a perfect matching $M'$ of
$G$. The set $\mathcal M'$ of perfect matchings so obtained is an
odd covering of $G$, a contradiction.
\end{prf}

\begin{prop}\label{prop:OddCoveringEtTau5}
Let $G_1^{x_1}$ and $G_2^{x_2}$ be cubic graphs with distinguished vertices $x_1$ and $x_2$ such that $\tau(G_i^{x_i})\geq 5$ $(i=1,2$) and $\tau_{odd}(G_i^{x_i})\neq 5$ ($i=1,2$). Let $G_4^{x'}$ and $G_3^{y'}$ be two copies of the cubic graph on two vertices and  $G=K_4[G_1^{x_1}, G_2^{x_2}, G_3^{y'}, G_4^{x'}]$, then $\tau(G)\geq 5$ and if $\tau_{odd}{G}$ is defined  then $\tau_{odd}(G)\neq 5$.
\end{prop}
\begin{prf}
Let $x$ and $y$ be respectively the unique vertex of $G_4$, $G_3$ (see Figure \ref{Fig:K4Composition} where $G_4$ is reduced to a single vertex $x$ and $G_3$ is reduced to $y$).
We know by Proposition \ref{Proposition:Constructiontau5} that$\tau(G)\geq 5$.
Assume that $\tau_{odd}(G)=5$ and let $\mathcal M=\{M_1, M_2, M_3, M_4, M_5\}$ be an odd $5$-covering. The perfect matchings of $\mathcal M$ are pairwise distinct otherwise by Proposition \ref{Proposition:CouvertureImpairePerfectMatchingsDisjoint} either $G_1^{x_1}$ or $G_2^{x_2}$ would be $3$-edge colorable, a contradiction. Observe that each vertex is incident to one edge that belongs to precisely three matchings of $\mathcal M$, the two other edges being covered only once. Moreover, the set of edges that belong to $3$ matchings of $\mathcal M$ is a perfect matching itself.

\vspace{\baselineskip}The $3$-edge cut $\{a_1y, b_1x, c_1c_2\}$ must be entirely contained in some matching of $\mathcal M$, say $M_i$ otherwise we would have a $5$- odd covering of $G_1^{x_1}$, a contradiction. Similarly there is a perfect matching in $\mathcal M$, say $M_j$ that contains the edges $c_1c_2$, $b_2x$, $a_2y$. Thus the edge $c_1c_2$ must belong to $3$ matchings of $\mathcal M$. Without loss of generality we assume that $i=1$, $j=2$ and $c_1c_2\in M_1\cap M_2\cap M_3$.

\vspace{\baselineskip}If $ya_1\in M_3$, since a perfect matching intersects any odd cut in an odd number of edges we have $xb_1\in M_3$, it follows that the edge $ya_1$ must be a member of a third matching of $\mathcal M$ as well as the edge $xb_1$. If for some $k$ we have $ya_1\in M_k$ and $xb_1\in M_k$, $k\in\{2,4,5\}$, $k$ being obviously distinct from $2$ $M_k$ intersects the $3$-edge cut in an even number of edges, a contradiction. Hence we may assume that $ya_1\in M_4$ and $xb_1\in M_5$. But now the edge $xy$ is covered by none of the matchings of $\mathcal M$, a contradiction. Consequently $ya_1\notin M_3$, similarly $xb_1\notin M_3$.

\vspace{\baselineskip}If $ya_1\in M_4$ this edge must belong to a third matching of $\mathcal M$ which is $M_5$. Since the set of edges that are covered $3$ times is a perfect matching $xb_1\in M_4\cap M_5$. But in this case the edge $c_1c_2$ would belong to $M_4$ and $M_5$, a contradiction.

\vspace{\baselineskip}It follows that $ya_1$ as well as $xb_2$ are covered only once and the edge $xy$ belongs to $3$ matchings of $\mathcal M$, that is $xy\in M_3\cap M_4\cap M_5$. But now, neither $M_4$ nor $M_5$ intersect the edge-cut $\{ya_1, xb_1,c_1c_2\}$ a contradiction since a perfect matching must intersect every odd edge-cut in an odd number of edges.
\end{prf}

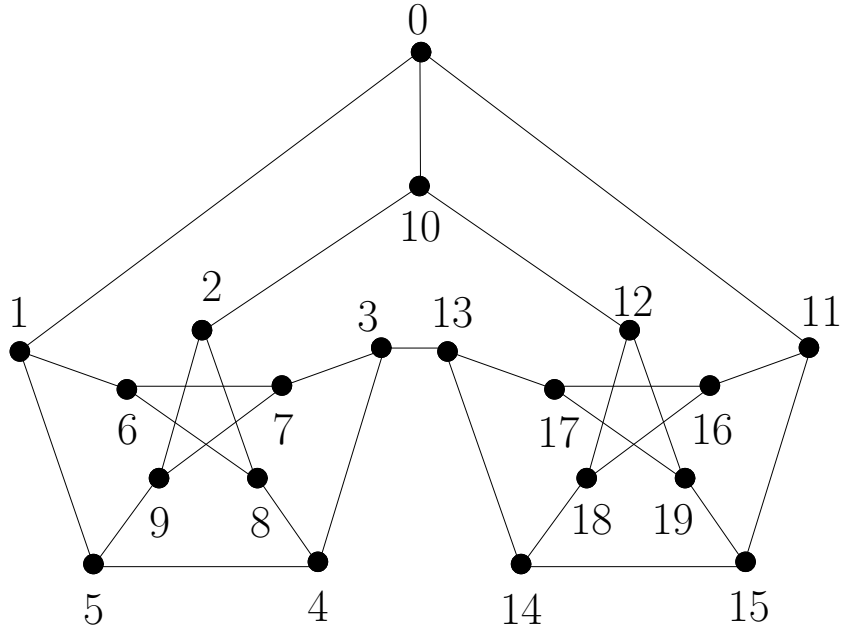
\begin{figure}[htb]
\begin{center}
\setlength{\unitlength}{0.00030621in}
\begingroup\makeatletter\ifx\SetFigFont\undefined
% extract first six characters in \fmtname
\def\x#1#2#3#4#5#6#7\relax{\def\x{#1#2#3#4#5#6}}%
\expandafter\x\fmtname xxxxxx\relax \def\y{splain}%
\ifx\x\y   % LaTeX or SliTeX?
\gdef\SetFigFont#1#2#3{%
  \ifnum #1<17\tiny\else \ifnum #1<20\small\else
  \ifnum #1<24\normalsize\else \ifnum #1<29\large\else
  \ifnum #1<34\Large\else \ifnum #1<41\LARGE\else
     \huge\fi\fi\fi\fi\fi\fi
  \csname #3\endcsname}%
\else
\gdef\SetFigFont#1#2#3{\begingroup
  \count@#1\relax \ifnum 25<\count@\count@25\fi
  \def\x{\endgroup\@setsize\SetFigFont{#2pt}}%
  \expandafter\x
    \csname \romannumeral\the\count@ pt\expandafter\endcsname
    \csname @\romannumeral\the\count@ pt\endcsname
  \csname #3\endcsname}%
\fi
\fi\endgroup
{\renewcommand{\dashlinestretch}{30}
\begin{picture}(14407,11141)(0,-10)
\put(6704,6563){\makebox(0,0)[lb]{\smash{{{\SetFigFont{17}{20.4}{rm}10}}}}}
\put(7535,4673){\blacken\ellipse{336}{336}}
\put(7535,4673){\ellipse{336}{336}}
\put(13735,4736){\blacken\ellipse{336}{336}}
\put(13735,4736){\ellipse{336}{336}}
\put(7053,7511){\blacken\ellipse{336}{336}}
\put(7053,7511){\ellipse{336}{336}}
\put(9373,4023){\blacken\ellipse{336}{336}}
\put(9373,4023){\ellipse{336}{336}}
\put(12035,4086){\blacken\ellipse{336}{336}}
\put(12035,4086){\ellipse{336}{336}}
\put(7083,9810){\blacken\ellipse{336}{336}}
\put(7083,9810){\ellipse{336}{336}}
\put(6397,4736){\blacken\ellipse{336}{336}}
\put(6397,4736){\ellipse{336}{336}}
\put(9923,2498){\blacken\ellipse{336}{336}}
\put(9923,2498){\ellipse{336}{336}}
\put(11610,2498){\blacken\ellipse{336}{336}}
\put(11610,2498){\ellipse{336}{336}}
\put(12648,1061){\blacken\ellipse{336}{336}}
\put(12648,1061){\ellipse{336}{336}}
\put(8798,1023){\blacken\ellipse{336}{336}}
\put(8798,1023){\ellipse{336}{336}}
\put(5310,1061){\blacken\ellipse{336}{336}}
\put(5310,1061){\ellipse{336}{336}}
\put(1460,1023){\blacken\ellipse{336}{336}}
\put(1460,1023){\ellipse{336}{336}}
\put(2585,2498){\blacken\ellipse{336}{336}}
\put(2585,2498){\ellipse{336}{336}}
\put(4272,2498){\blacken\ellipse{336}{336}}
\put(4272,2498){\ellipse{336}{336}}
\put(2035,4023){\blacken\ellipse{336}{336}}
\put(2035,4023){\ellipse{336}{336}}
\put(4697,4086){\blacken\ellipse{336}{336}}
\put(4697,4086){\ellipse{336}{336}}
\put(3322,5036){\blacken\ellipse{336}{336}}
\put(3322,5036){\ellipse{336}{336}}
\put(197,4673){\blacken\ellipse{336}{336}}
\put(197,4673){\ellipse{336}{336}}
\path(13788,4734)(7068,9836)
\path(187,4674)(7061,9827)
\path(7068,7519)(7061,9827)
\path(6409,4729)(7525,4729)(7525,4747)
\path(12613,996)(13766,4729)
\path(8788,977)(12613,977)
\path(7489,4693)(8788,977)
\path(5311,977)(6446,4711)
\path(1468,977)(5311,977)
\path(187,4711)(1468,977)
\path(9910,2511)(8823,1024)
\path(7535,4674)(9398,4011)
\path(11585,2474)(10673,5074)
\path(9310,4074)(11585,2474)
\path(12060,4074)(9310,4074)
\path(9923,2474)(12060,4074)
\path(10648,5036)(9923,2474)
\path(11598,2524)(12623,1049)
\path(12035,4074)(13773,4699)
\path(7023,7519)(10660,5011)
\path(2572,2511)(1485,1024)
\path(197,4674)(2060,4011)
\path(4247,2474)(3335,5074)
\path(1972,4074)(4247,2474)
\path(4722,4074)(1972,4074)
\path(2585,2474)(4722,4074)
\path(3310,5036)(2585,2474)
\path(4260,2524)(5285,1049)
\path(4697,4074)(6435,4699)
\path(7068,7519)(3322,5011)
\put(4147,1490){\makebox(0,0)[lb]{\smash{{{\SetFigFont{17}{20.4}{rm}8}}}}}
\put(2416,1490){\makebox(0,0)[lb]{\smash{{{\SetFigFont{17}{20.4}{rm}9}}}}}
\put(1288,0){\makebox(0,0)[lb]{\smash{{{\SetFigFont{17}{20.4}{rm}5}}}}}
\put(5133,40){\makebox(0,0)[lb]{\smash{{{\SetFigFont{17}{20.4}{rm}4}}}}}
\put(8435,0){\makebox(0,0)[lb]{\smash{{{\SetFigFont{17}{20.4}{rm}14}}}}}
\put(12340,40){\makebox(0,0)[lb]{\smash{{{\SetFigFont{17}{20.4}{rm}15}}}}}
\put(11044,1530){\makebox(0,0)[lb]{\smash{{{\SetFigFont{17}{20.4}{rm}19}}}}}
\put(9643,1530){\makebox(0,0)[lb]{\smash{{{\SetFigFont{17}{20.4}{rm}18}}}}}
\put(11716,3080){\makebox(0,0)[lb]{\smash{{{\SetFigFont{17}{20.4}{rm}16}}}}}
\put(9079,3040){\makebox(0,0)[lb]{\smash{{{\SetFigFont{17}{20.4}{rm}17}}}}}
\put(4529,3080){\makebox(0,0)[lb]{\smash{{{\SetFigFont{17}{20.4}{rm}7}}}}}
\put(1872,3080){\makebox(0,0)[lb]{\smash{{{\SetFigFont{17}{20.4}{rm}6}}}}}
\put(13587,5112){\makebox(0,0)[lb]{\smash{{{\SetFigFont{17}{20.4}{rm}11}}}}}
\put(3311,5545){\makebox(0,0)[lb]{\smash{{{\SetFigFont{17}{20.4}{rm}2}}}}}
\put(10347,5275){\makebox(0,0)[lb]{\smash{{{\SetFigFont{17}{20.4}{rm}12}}}}}
\put(7257,5092){\makebox(0,0)[lb]{\smash{{{\SetFigFont{17}{20.4}{rm}13}}}}}
\put(5991,5022){\makebox(0,0)[lb]{\smash{{{\SetFigFont{17}{20.4}{rm}3}}}}}
\put(0,5092){\makebox(0,0)[lb]{\smash{{{\SetFigFont{17}{20.4}{rm}1}}}}}
\put(6845,10126){\makebox(0,0)[lb]{\smash{{{\SetFigFont{17}{20.4}{rm}0}}}}}
\put(10660,5036){\blacken\ellipse{336}{336}}
\put(10660,5036){\ellipse{336}{336}}
\end{picture}
}
\caption{A graph $G$ such that $\tau(G)=5$ and $\tau_{odd}(G)=7$.}
\label{Fig:Tau5Odd}
\end{center}
\end{figure}

The graph $G$ depicted in  Figure \ref{Fig:Tau5Odd} is an example of cubic graphs with a $7$-odd covering and a perfect matching index equals to $5$. We know by Proposition \ref{prop:OddCoveringEtTau5} that $\tau_{odd}(G)\geq 7$. As a matter of fact, this graph has $20$ distinct perfect matchings and among all the $7$-tuples of perfect matchings ($77520$) $64$ form an odd-covering. Let us give below such a $7$-tuple.
\begin{center}
\begin{eqnarray*}
\{0-10,\;1-5,\;2-9,\;3-13,\; 4-8,\;6-7,\;11-15,\;12-19,\;14-18,\;16-17\}\\
\{ 0-1,\;2-8,\;3-4,\;5-9,\;6-7,\;10-12,\;11-15,\;13-14,\;16-18,\;17-19\}\\
\{ 0-1,\;2-10,\;3-13,\;4-5,\;6-8,\;7-9,\;11-15,\;12-19,\;14-18,\;16-17\}\\
\{ 0-1,\;2-10,\;3-13,\;4-8,\;5-9,\;6-7,\;11-16,\;12-18,\;14-15,\;17-19\}\\
\{ 0-11,\;1-5,\;2-9,\;3-13,\;4-8,\;6-7,\;10-12,\;14-15,\;16-18,\;17-19\}\\
\{ 0-11,\;1-5,\;2-9,\;3-13,\;4-8,\;6-7,\;10-12,\;14-18,\;15-19,\;16-17\}\\
\{ 0-11,\;1-6,\;2-10,\;3-7,\;4-8,\;5-9,\;12-19,\;13-17,\;14-15,\;16-18\}\\
\end{eqnarray*}
\end{center}
Moreover the following perfect matchings form a $5$-covering.
\begin{center}
\begin{eqnarray*}
\{0-1,\;2-10,\;3-13,\;6-8,\;7-9,\;4-5,\;12-19,\;16-17,\;14-18,\;11-15\}\\
\{2-9,\;1-6,\;7-9,\;4-5,\;3-13,\;0-11,\;10-12,\;14-15,\;16-18,\;17-19\}\\
\{1-6,\;7-9,\;2-8,\;5-4,\;0-10,\;12-18,\;17-19,\;14-15,\;11-15,\;3-13\}\\
\{0-1,\;2-8,\;6-7,\;5-9,\;3-4,\;10-12,\;13-17,\;14-18,\;15-19,\;11-16\}\\
\{1-6,\;5-9,\;4-8,\;3-7,\;2-10,\;0-11,\;12-18,\;13-14,\;15-19,\;16-17\}\\
\end{eqnarray*}
\end{center}

We do not know any example of graph $G$ for which $\tau_{odd}$ is defined and with $\tau(G)=\tau_{odd}(G)=5$. We just observe that in such a graph every vertex would be incident to an edge belonging  to $3$ perfect matchings and to precisely two edges covered only once. The set of edges covered by $3$ perfect matchings being a perfect matching itself.

\begin{prob}\label{Problem:CouverturePaireSize2ou4} Is it true that every bridgeless cubic graph has an even covering where each edge appears twice or $4$ times~?
\end{prob}

The answer is yes for $3-$edge-colorable cubic graphs
and for bridgeless cubic graphs with perfect matching index $4$
since such graphs  have an even covering of size $8$.
% ----------------------------------------------------------------
\bibliographystyle{amsplain}
\bibliography{BibliographieBergeFulkerson}
\end{document}